\documentclass[12pt]{article}
\usepackage{float}
\usepackage[utf8]{inputenc} 
\usepackage{amsmath, amssymb} 
\usepackage{graphicx}  
\usepackage[margin=1in]{geometry}
\usepackage{tcolorbox}
\usepackage{natbib}
\usepackage[left]{lineno}
\linenumbers
\usepackage{rotating}
\usepackage{multirow}
\usepackage{booktabs}
\usepackage{multirow}
\usepackage{color}
\usepackage{hyperref}
\usepackage{chngcntr}
\usepackage{makecell}

\title{A neural operator framework for data-driven discovery of stability and receptivity in physical systems}

\author{
    Chengyun Wang$^{1}$, Liwei Chen$^{2}$ \& Nils Thuerey$^{1,*}$ \\
    \\
    {\small $^{1}$Department of Computer Science, School of Computation, Information and Technology,} \\
    {\small Technical University of Munich, Boltzmannstrasse 3, D-85748 Garching bei München, Germany} \\
    {\small $^{2}$Beijing Institute of Astronautical Systems Engineering, Beijing, China} \\
    {\small $^{*}$Corresponding author: nils.thuerey@tum.de}
}

\date{} 

\begin{document}
\nolinenumbers
\maketitle

\begin{tcolorbox}[colback=blue!10,title=Significance]
Investigating how complex systems respond to perturbations remains a central challenge in both science and engineering. Traditional analysis methods are limited to simple, linearized systems or require explicit governing equations. We introduce a general, data-driven framework that uses neural networks to discover the stability and receptivity properties of nonlinear systems from trajectory data. Our approach identifies which perturbations will grow unstably around both equilibrium points and periodic orbits and crucially pinpoints the optimal way to force the system to elicit the most amplified responses, even for strongly nonlinear systems. This framework is relevant to applications in fields ranging from climate science and neuroscience to fluid engineering, particularly where first-principles models are often intractable but sufficiently rich trajectory data is available.
\end{tcolorbox}

\section*{Abstract}
Understanding how complex systems respond to perturbations, such as whether they will remain stable or what their most sensitive patterns are, is a fundamental challenge across science and engineering. Traditional stability and receptivity (resolvent) analyses are powerful but rely on known equations and linearization, limiting their use in nonlinear or poorly modeled systems. Here, we introduce a data-driven framework that automatically identifies stability properties of both equilibrium points and periodic orbits, as well as optimal forcing responses from observation data alone, without requiring governing equations. By training a neural network as a dynamics emulator and using automatic differentiation to extract its Jacobian, we can perform stability and receptivity analyses directly from data. We demonstrate the method on systems of increasing complexity, from canonical models to high‑dimensional fluid flows and a real‑world climate dataset, employing model reduction where needed and successfully identifying dominant instability modes and optimal input–output structures, even in strongly nonlinear regimes. By leveraging a neural network emulator, we readily obtain a nonlinear representation of system dynamics while additionally retrieving intricate dynamical patterns that were previously difficult to resolve. This equation-free methodology establishes a broadly applicable tool for analyzing complex, high-dimensional datasets, with potential relevance to grand challenges in fields such as climate science, neuroscience, and fluid engineering.

\section{Introduction}
Real-world systems are inherently nonlinear and exhibit rich, multi-scale behavior in both space and time. Dynamical system theory offers a unified mathematical framework to describe, analyze, and predict the evolution of such systems, capturing the complex interactions among quantities that evolve over time. From its foundations in the seminal work of Poincaré on celestial mechanics, this field has expanded to encompass a vast range of phenomena across the engineering, physical, and life sciences. Within this universal framework, a fundamental challenge persists: to understand and predict how a system responds to perturbations. Answering this involves determining if a system's state is stable, identifying the conditions that trigger transitions to instability, and discovering which external inputs will most effectively steer the system's evolution.

Given that nonlinearity remains a primary challenge in analyzing and controlling dynamical systems, a leading perspective is to investigate the system's local behavior through linearization to extract the dominant modes of perturbation evolution \citep{guckenheimer2013nonlinear}. This strategy has given rise to a powerful suite of modal analysis techniques, many of which were developed within the hydrodynamics and cybernetics community \citep{simon1968theory, barkley1996three, holmes2012turbulence, proctor2016dynamic, rowley2017model}. These techniques can be broadly divided into two categories \citep{taira2017modal}. The first consists of \textit{operator-based methods}, such as stability analysis \citep{theofilis2003advances,barkley1996three, robichaux1999three,theofilis2011global}, and resolvent analysis \citep{trefethen1993hydrodynamic,mckeon2010critical}, which are built upon the linearized operator derived from the system's governing equations. Their profound insights, however, are contingent on this explicit, model-based knowledge. To address systems not amenable to first principles, a second category of \textit{data-driven methods} has emerged. These approaches, including dynamic mode decomposition (DMD) \citep{schmid2010dynamic,kutz2016dynamic}, proper orthogonal decomposition (POD) \citep{lumley1967structure,holmes2012turbulence}, and sparse identification of nonlinear dynamics (SINDy)\citep{brunton2016discovering}, rely purely on observational data to uncover underlying dynamical structures, offering a pathway to analyze systems where governing equations are unknown.

Stability analysis and resolvent analysis, as two representative operator-based methods, have been demonstrated to be highly effective in providing physical insights into the stability and receptivity properties of a broad range of dynamical systems. The former can be further divided into \textit{linear stability analysis} which examines the neighborhood of a fixed point, and \textit{Floquet analysis} which examines the neighborhood of a periodic orbit. Both rest on the same idea: they linearize the governing equations around the base state and assume exponential perturbation growth. This assumption elegantly converts the linear initial-value problem into an eigenvalue problem. The stability of a fixed point is governed by the eigenvalues of the Jacobian matrix, while the stability of a periodic orbit is governed by the eigenvalues of the monodromy matrix that propagates perturbations over one period. However, this framework is limited to describing the asymptotic perturbation evolution and does not account for short-term transient behavior \citep{schmid2007nonmodal}.

To remedy this shortcoming, resolvent analysis was developed to investigate the receptivity of the linearized system to external forcings. Instead of focusing on internal instabilities, it can identify the optimal forcing patterns that elicit the most amplified responses, which usually represent the natural starting point for control design \citep{jovanovic2021bypass}. As an input-output response framework, it provides not only a transfer-function viewpoint of analyzing the system's receptivity characteristics, but also reveals the transient growth of perturbations, thereby explaining the transient energy amplification that arises in linearly stable systems. Therefore, stability analysis and resolvent analysis complement each other to provide a comprehensive picture of the system's linearized dynamics.

Despite the effectiveness of operator-based methods, their practical application to many dynamical systems of interest remains very challenging. In numerous scientific frontiers, such as neuroscience, epidemiology, and ecology, there is a fundamental lack of known physical laws from which to derive the governing equations of motion. Even in systems where such equations are well-established, such as turbulence and combustion, the construction of the linearized operator often requires specialized numerical solvers or adjoint simulations \citep{rolandi2024invitation}. Meanwhile, the rapid progress in experimental measurements and numerical simulations has led to data generation at an unprecedented scale. This data deluge is driving a paradigm shift to data-driven methods that can extract meaningful information directly from datasets without prior knowledge. 

Therefore, these two classical categories are being complemented by a promising hybrid perspective that seeks to perform operator-based analysis in a purely data-driven manner. A leading example is a DMD-based approach to resolvent analysis for linear or weakly nonlinear systems, in which DMD is applied to time-resolved flow snapshots to construct the resolvent operator \citep{herrmann2021data}. This idea is further refined by incorporating physics-informed constraints to restrict the admissible DMD model space to certain matrix manifolds that preserve the desired physical properties \citep{baddoo2023physics}. However, the linear assumption of DMD becomes increasingly restrictive as nonlinear effects grow. A notable recent improvement is the linear and nonlinear disambiguation optimization (LANDO) framework \citep{baddoo2022kernel}, which uses kernel learning to identify nonlinear dynamics while isolating an interpretable linear component, and has already shown promising results in hydrodynamic stability applications \citep{gomez2023toward}.

Neural networks (NNs) offer a pathway beyond these limitations. As universal function approximators, NN emulators can learn complex nonlinear dynamics from spatiotemporal measurements of the system, serving as surrogate models that capture both linear and nonlinear behaviors \citep{floryan2022data,yu2024learning,page2024recurrent}. They can be trained either in a purely data-driven manner \citep{morton2018deep,thuerey2020,chen2023towards} or augmented with physical information \citep{barsinai2019data,raissi2019physics,kochkov2021machine,list2022learned,chen2024deep}. 
Most closely related to the present work, Déda \textit{et al.}~\citep{deda2023backpropagation,deda2024neural} exploit Jacobians of an input-augmented neural map to locate equilibria 
and build control-oriented linear models. However, they compute the Jacobians at the fixed points only for pole‑placement stabilization, and gather locally linear data around those points to conduct data-driven stability analysis via DMD with control, while resolvent analysis is left as an open direction. These are precisely the gaps our approach closes: the local Jacobian of the NN emulator can be extracted at any point along the data trajectory via automatic differentiation \citep{chen2024deep}, opening the door to an equation-free, data-driven modal analysis for nonlinear systems.

To this end, we propose a data-driven approach to learn a NN emulator from data generated by truly nonlinear systems, with a schematic illustration presented in Fig.~\ref{fig:schematic}. The Jacobian of the trained NN emulator serves as an approximation of the system's local linear operator, which is subsequently used to perform operator-based modal analysis. To demonstrate the power and generality of this framework, we diagnose four representative dynamical systems and successfully extract their essential stability characteristics and input-output resolvent modes. For high-dimensional systems, we additionally introduce a reduced-order formulation that makes the learning process more efficient while retaining the leading spectral information. Furthermore, by comparing with traditional data-driven methods like DMD, we provide insights on how to select the appropriate tool for data-driven analysis.

\section{Methods}\label{sec:method2}

Considering a nonlinear dynamical system (temporally continuous but spatially discretized into $N$ degrees of freedom), the governing equation can be expressed in terms of the state variable $\boldsymbol{q} \in \mathbb{C}^N$ in the compact operator form as

\begin{equation}\label{equ:dynamic}
\dot{\boldsymbol{q}} = \mathcal{N}(\boldsymbol{q}),
\end{equation}

\noindent where $\mathcal{N}$ is the nonlinear operator governing the dynamical behavior of the system. To linearize this ordinary differential equation, the state variable can be decomposed as a sum of base state $\boldsymbol{q}_b \in \mathbb{C}^N$ and infinitesimal perturbation $\boldsymbol{q}' \in \mathbb{C}^N$, which yields

\begin{equation}\label{equ:lineardynamic1}
\dot{\boldsymbol{q}_b} + \dot{\boldsymbol{q}}' = \mathcal{N}(\boldsymbol{q}_b+\boldsymbol{q}') = \mathcal{N}(\boldsymbol{q}_b) + \frac{ \partial \mathcal{N}}{\partial \boldsymbol{q}}\Big|_{\boldsymbol{q}_b } \boldsymbol{q}' + O(|\boldsymbol{q}'|^2).
\end{equation}

The base state can be any solution satisfying $d \boldsymbol{q}_b/d t = \mathcal{N}(\boldsymbol{q}_b)$, so that we
can eliminate both terms and 
neglect the higher-order infinitesimals in Eq.~\ref{equ:lineardynamic1}. Then we obtain the linearized equation of the perturbation $\boldsymbol{q}'$ as 

\begin{equation}\label{equ:lineardynamic2}
\dot{\boldsymbol{q}}' = \frac{\partial \mathcal{N}}{\partial \boldsymbol{q}}\Big|_{\boldsymbol{q}_b}\boldsymbol{q}' = \mathbf{A}\boldsymbol{q}',
\end{equation}

\noindent where $\mathbf A \in \mathbb{C}^{N \times N}$ represents the discretized linear operator of this continuous dynamical system at base state $\boldsymbol{q}_b$, i.e., the local Jacobian. 

Based on the local Jacobian, we can evaluate the stability around fixed points satisfying $d \boldsymbol{q}_b/d t =0$ (linear stability analysis), or even periodic orbits where Jacobian $\mathbf{A}(t)$ is periodic (Floquet analysis). We can also assess the system's receptivity by including an additional forcing term $\boldsymbol{f}$ to identify the optimal forcing frequency/mode (resolvent analysis). Details about these operator-based modal analyses are provided in \textit{SI Appendix}, Sec.~1.

\subsection{Neural network operator}\label{sec:nnop}

\begin{figure}[t!]
    \centering
    \includegraphics[width=0.5\textwidth]
    {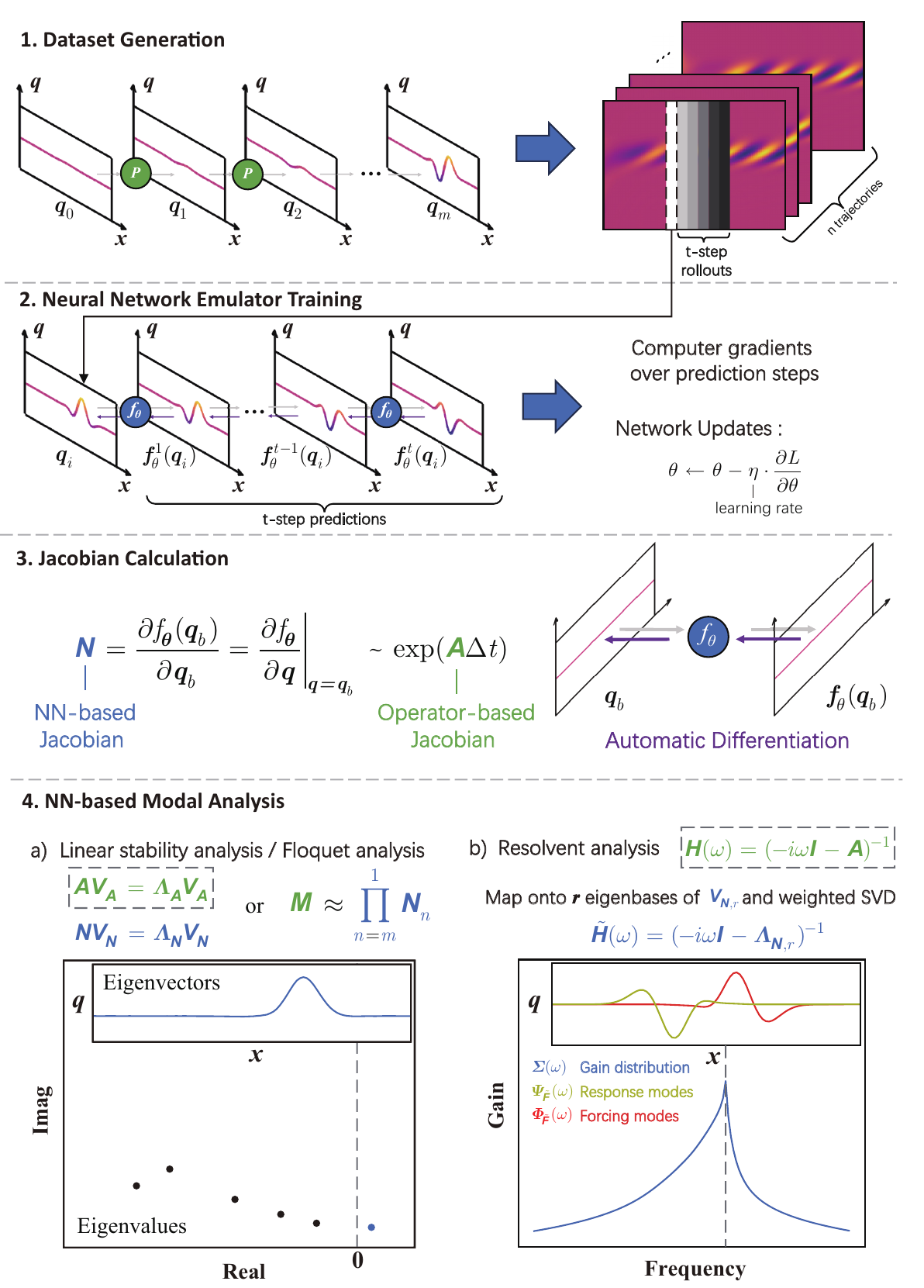}
    \caption{Schematic of our data-driven algorithm. (1) Time-resolved state snapshots collected from a dynamical system are used. (2) An NN emulator is trained using a rollout strategy to accurately learn the system's evolution map. (3) The Jacobian of the trained emulator is then extracted via automatic differentiation and used to approximate the local Jacobian. This data-driven operator is the key ingredient for (4) NN-based modal analysis to perform linear stability analysis or Floquet analysis via an eigenvalue decomposition to find unstable modes, and resolvent analysis via a weighted SVD on the projected resolvent to identify optimal forcing modes, response modes, and the gain distribution.}
    \label{fig:schematic}
\end{figure}

We adopt a time-stepping framework using a NN emulator to evolve the dynamical system of Eq.~\ref{equ:dynamic}. The state $\boldsymbol{q}_{n}$ at time $t$ is mapped to the subsequent state $\boldsymbol{q}_{n+1}$ over a time increment $\Delta t$ as:

\begin{equation}\label{equ:disdynamic}
\boldsymbol{q}_{n + 1} = f_\theta ( \boldsymbol{q}_n),
\end{equation}

\noindent where $f_\theta$ is an arbitrary NN architecture with weights $\theta$. Let $P$ represent a classical numerical solver that also advances the system state with a time step $\Delta t$. Then the learning task is to find the optimal weights $\theta_{\text{opt}}$ of the NN to best approximate the numerical solver, i.e., $f_\theta \approx  P$. When trained on a sufficiently diverse dataset, the NN emulator can capture the underlying latent dynamics and generalize effectively to previously unseen system states.

Since the network is trained by minimizing the difference between predicted and true trajectories, their interplay during the learning process is crucial. A straightforward way of training is one-step training, which only minimizes the one-step prediction error \citep{tran2021factorized}. By contrast, rollout training by autoregressively applying $f_\theta$ for multiple time steps can learn long-term dependencies and correct for compounding errors \citep{um2020solver}. This can significantly improve the long-term prediction performance, especially when dealing with high-dimensional nonlinear dynamical systems. 

Suppose the ground truth trajectories produced by $P$ are ${\boldsymbol{q}}^{(j)}_0, {\boldsymbol{q}}^{(j)}_1, \cdots, {\boldsymbol{q}}^{(j)}_m$, where the superscript $j \in \left\{1, \cdots, n\right\}$ indexes the trajectories corresponding to $n$ distinct initial conditions, and $m$ is the maximum sampling time step per trajectory (Part 1, Fig.~\ref{fig:schematic}). Given an arbitrary state ${\boldsymbol{q}}^{(j)}_i$ extracted from the ground truth datasets, the trajectories predicted by iteratively applying the learned model $f_\theta$ for $t$ steps are denoted as ${\boldsymbol{q}}^{(j)}_i, f^1_\theta({\boldsymbol{q}}^{(j)}_i), \cdots, f^t_\theta({\boldsymbol{q}}^{(j)}_i)$. The total loss function for training with $t$-step rollouts averaged on $n$ trajectories is:

\begin{equation}
    \mathcal{L}(\theta) = \frac{1}{n} \sum_{j=1}^{n} \sum_{i=0}^{m-t} \sum_{k=1}^{t} \left[ \mathcal{L}_2 \left(f_\theta^k({\boldsymbol{q}}^{(j)}_i), {\boldsymbol{q}}^{(j)}_{i+k} \right) \right],
\end{equation}

\noindent where $\mathcal{L}_2$ represents L2-norm loss function. Full gradients can be computed owing to the differentiability of the NN emulator, which enables backpropagation through $t$ rollout steps and subsequent updates of the network weights $\theta$ (Part 2, Fig.~\ref{fig:schematic}).

\subsection{Relationship between NN-based and operator-based Jacobians} Since the NN emulator operates in a temporally discrete manner, whereas operator-based analysis relies on temporally continuous equations, an essential step is to bridge the gap between these two representations of the dynamical system. 

Here, we can also linearize the temporally discrete dynamical system (Eq.~\ref{equ:disdynamic}) based on the base-perturbation decomposition:

\begin{equation}\label{equ:dis_perturbation}
\boldsymbol{q}_{n+1} + \boldsymbol{q}'_{n+1} =
f_\theta(\boldsymbol{q}_n + \boldsymbol{q}'_n) = f_\theta(\boldsymbol{q}_n) + \frac{\partial f_\theta}{\partial \boldsymbol{q}}\Big|_{\boldsymbol{q}_n}\boldsymbol{q}'_n + O(|\boldsymbol{q}'_n|^2).
\end{equation}

Since $\boldsymbol{q}_{n}$ and $\boldsymbol{q}_{n+1}$ are two successive states on the trajectory governed by the discrete dynamics Eq.~\ref{equ:disdynamic}, we can eliminate both terms in Eq.~\ref{equ:dis_perturbation}, and neglect the high-order term to obtain:

\begin{equation}
\boldsymbol{q}'_{n+1} = \frac{\partial f_\theta}{\partial \boldsymbol{q}}\Big|_{\boldsymbol{q}_n} \boldsymbol{q}'_n = \mathbf{N} \boldsymbol{q}'_n,
\end{equation}

\noindent where the Jacobian $\mathbf{N}$ of our learned neural operator $f_\theta$ can be calculated via automatic differentiation evaluated at $\boldsymbol{q}_n$ (Part 3, Fig.~\ref{fig:schematic}). The differentiability of the NN emulator actually allows for the use of automatic differentiation to evaluate $\mathbf{N}$ at any input state.

The general solution of Eq.~\ref{equ:lineardynamic2} takes an exponential form, and thus two successive perturbation states $\boldsymbol{q}'$ should satisfy:

\begin{equation}
\boldsymbol{q}'_{n+1} = \exp (\mathbf{A} \Delta t) \boldsymbol{q}'_n.
\end{equation}

Hence, we can derive the relationship between the NN-based Jacobian $\mathbf{N}$ and the operator-based Jacobian $\mathbf{A}$ as

\begin{equation}\label{equ:relation}
\mathbf{N} = \exp (\mathbf{A} \Delta t).
\end{equation}

Benefiting from this relationship, if a learned surrogate model can accurately capture the underlying system dynamics, it can then be utilized to evaluate the stability of the original system without knowing its governing equations. 

\subsection{NN-based stability and receptivity analysis}\label{sec:3.3}

For linear stability analysis, we are concerned only with the eigenvalue information of $\mathbf{A}$, which determines whether a perturbation about the base state $\boldsymbol{q}_b$ will grow. According to Eq.~\ref{equ:relation}, the eigenvalues of the NN-based and operator-based Jacobians, denoted by the diagonal matrices $\boldsymbol{\Lambda}_{\mathbf{N}}$ and $\boldsymbol{\Lambda}_{\mathbf{A}}$, respectively, are expected to satisfy the exponential relationship $\boldsymbol{\Lambda}_{\mathbf{N}} = \exp(\boldsymbol{\Lambda}_{\mathbf{A}} \Delta t)$, while the eigenvectors associated with each eigenvalue should be identical, i.e., $\mathbf{V}_\mathbf{N} = \mathbf{V}_\mathbf{A}$.

Provided the learned model accurately approximates the original system, the eigenvalues and eigenvectors extracted from $\mathbf{N}$ serve as good approximations of those of $\mathbf{A}$, whose spectral characteristics can thereby be inferred indirectly (Part~4a, Fig.~\ref{fig:schematic}). This constitutes the core procedure of NN-based linear stability analysis. 

For Floquet analysis, the quantity of interest is the monodromy matrix $\mathbf{M}$, which propagates the perturbation by one period $T$ according to $\boldsymbol{q}'(t+T)=\mathbf{M}\boldsymbol{q}'(t)$. It can be regarded as the periodic-orbit counterpart of $\mathbf{N}$ in the linear stability analysis. We sample the periodic base state  over one period at $m$ discrete instants $\boldsymbol{q}_b(t_1),\dots,\boldsymbol{q}_b(t_m)$, with $t_{n+1}=t_n+\Delta t$ and $T=m\Delta t$. Then the local one-step propagator $\mathbf{G}_n$ that advances the perturbation from $t_n$ to $t_{n+1}$ can be approximated by the NN-based Jacobian $\mathbf{N}_n$ evaluated at each instant:

\begin{equation}\label{equ:monodromy_nn} 
\mathbf{G}_n = \mathcal{T}\exp \int_{t_n}^{t_n+\Delta t}\mathbf{A}(\tau)\,\mathrm{d}\tau \approx \mathbf{N}_n=\frac{\partial f_\theta}{\partial \boldsymbol{q}}\bigg|_{\boldsymbol{q}_b(t_n)},
\end{equation}

\noindent where $\mathcal{T}\exp$ denotes the time-ordered exponential. The monodromy matrix is assembled from the ordered product of the local Jacobians along the orbit over one period,

\begin{equation}\label{equ:monodromy_product}
\mathbf{M}=\mathcal{T}\exp \int_{t}^{t+T}\mathbf{A}(\tau)\,\mathrm{d}\tau \approx \mathbf{N}_m\,\mathbf{N}_{m-1}\cdots\mathbf{N}_2\,\mathbf{N}_1=\prod_{n=m}^1 \mathbf{N}_n,
\end{equation}

\noindent which generalizes the fixed-point relation $\mathbf{N}=\exp(\mathbf{A}\Delta t)$ to a periodic base state. Its eigendecomposition yields the Floquet multipliers (eigenvalues) and modes (eigenvectors) that characterize the orbit’s stability (Part~4a, Fig.~\ref{fig:schematic}). Note that the orbit's starting point is immaterial, since monodromy matrices assembled from different starting phases are similar to one another, related by the propagator between the two instants. The Floquet multipliers are therefore independent of the starting point, while the corresponding modes are simply transported along the orbit by the same propagator.

For resolvent analysis, a natural idea is to replace the Jacobian $\mathbf{A}$ by $\log(\mathbf{N})/\Delta t$ in order to compute the resolvent operator $\mathbf{H}(\omega)=(-i\omega\mathbf{I}-\mathbf{A})^{-1}$. However, taking the logarithm of $\mathbf{N}$ directly is not advisable: the complex logarithm is multivalued and therefore strongly underdetermined, so that a given $\mathbf{N}$ corresponds to many matrices $\mathbf{A}$ satisfying Eq.~\ref{equ:relation}.

Instead, we can leverage the results of the linear stability analysis to implement the subspace projection \citep{reddy1993energy}. In a nutshell, we can project the forced system Eq.~\ref{equ:lineardynamic2} into the subspace expanded by a truncated $r$ eigenvectors \citep{schmid2012stability,herrmann2021data}. Supposing the original state variable $\boldsymbol{q}'$ and $\boldsymbol{f}$ in Eq.~\ref{equ:lineardynamic2} can be expressed in the subspace spanned by the first $r$ linearly independent eigenvectors $\mathbf{V}_{\mathbf{A},r}$, we have

\begin{equation}
\dot{\mathbf{V}}_{\mathbf{A},r}\boldsymbol{x} = \mathbf{A} \mathbf{V}_{\mathbf{A},r} \boldsymbol{x} + \mathbf{V}_{\mathbf{A},r} \boldsymbol{y},
\end{equation}

\noindent where $\boldsymbol{x}=[x_1,x_2,\cdots,x_r]^T \in \mathbb{C}^r$ and $\boldsymbol{y}=[y_1,y_2,\cdots,y_r]^T \in \mathbb{C}^r$ are the vectors of expansion coefficients in the eigenvector coordinate. Based on the definition of eigenvalue $\mathbf{A} \mathbf{V}_{\mathbf{A},r} = \mathbf{V}_{\mathbf{A},r}\boldsymbol{\Lambda}_{\mathbf{A},r}$, we can obtain the new governing equation in the eigenvector coordinate:

\begin{equation}
\dot{\boldsymbol{x}} = \boldsymbol{\Lambda}_{\mathbf{A},r} \boldsymbol{x} + \boldsymbol{y}.
\end{equation}

Given the weighting matrix $\mathbf{Q}$ used to define the state norm in the original coordinate as $|| \hat{\boldsymbol{q}} ||^2_{\mathbf{Q}}=\hat{\boldsymbol{q}}^*\mathbf{Q}\hat{\boldsymbol{q}}$, the corresponding weighting matrix in the new coordinate becomes $\tilde{\mathbf{Q}}= \mathbf{V}_{\mathbf{A},r}^* \mathbf{Q} \mathbf{V}_{\mathbf{A},r}$, which can be factorized as $\tilde{\mathbf{Q}} = \tilde{\mathbf{F}}^* \tilde{\mathbf{F}}$ via Cholesky factorization. Under this new coordinate system, we can carry out a weighted singular value decomposition (SVD) of the projected resolvent:

\begin{equation}
\tilde{\mathbf{F}} (-i\omega\mathbf{I}-\boldsymbol{\Lambda}_{\mathbf{A},r})^{-1} \tilde{\mathbf{F}}^{-1} = \boldsymbol{\Psi}_{\tilde{\mathbf{F}}}(\omega) \boldsymbol{\Sigma}(\omega) \boldsymbol{\Phi}^*_{\tilde{\mathbf{F}}}(\omega),
\end{equation}

\noindent where the diagonal entries of the singular value matrix $\Sigma(\omega)$ are the resolvent gains and the column vectors in $\boldsymbol{\Psi}_{\tilde{\mathbf{F}}}(\omega)$ and $\boldsymbol{\Phi}_{\tilde{\mathbf{F}}}(\omega) $ are the corresponding response and forcing modes at frequency $\omega$.
The conversion of resolvent modes back to the original space is then performed via 
$\boldsymbol{\Phi} = \mathbf{V}_{\mathbf{A},r} \tilde{\mathbf{F}}^{-1} \boldsymbol{\Phi}_{\tilde{\mathbf{F}}}(\omega)$ and $\boldsymbol{\Psi} = \mathbf{V}_{\mathbf{A},r} \tilde{\mathbf{F}}^{-1} \boldsymbol{\Psi}_{\tilde{\mathbf{F}}}(\omega)$. In practice, the derivation of $\boldsymbol{\Lambda}_{\mathbf{A},r}$ and $\mathbf{V}_{\mathbf{A},r}$ requires the knowledge of the corresponding governing equation, so $\boldsymbol{\Lambda}_{\mathbf{N},r}$ and $\mathbf{V}_{\mathbf{N},r}$ acquired from data are employed instead (Part 4b, Fig.~\ref{fig:schematic}). 

However, it is important to avoid indiscriminately increasing the dimensionality of the subspaces \citep{herrmann2021data}. A reliable projection of a high-dimensional system onto a reduced subspace requires that the chosen eigenvectors be linearly independent. But especially in non-normal systems, where eigenvectors are generally non-orthogonal, care must be taken not to select an overly large number of modes $r$. An excessively large $r$ may lead to a redundant coordinate system, resulting in a non-unique representation of the system state in terms of eigenvector coordinates. This will deteriorate the validity of resolvent operator estimation instead of enriching the basis. 

\subsection{Stability in reduced-order Jacobians}

For high-dimensional systems, it is natural to first represent the state snapshots in a POD basis and then learn the dynamics in the correspondingly smaller space of coefficients. 
Collecting $n$ snapshots from $m$ trajectories into a matrix $\mathbf{Q}\in\mathbb{C}^{N\times mn}$, and noting that typically $mn\ll N$, we employ the method of snapshots \citep{sirovich1987turbulence} to obtain the compact decomposition $\mathbf{Q}=\mathbf{U}\boldsymbol{\Sigma}\mathbf{W}^*$,
where $\mathbf{U}\in\mathbb{C}^{N\times l}$ has orthonormal columns and spans the POD subspace, with $l=\mathrm{rank}(\mathbf{Q})\le mn$. Hence, every snapshot $\boldsymbol{q}_n$ in the dataset can be converted as
\begin{equation}
\boldsymbol{a}_n=\mathbf{U}^*\boldsymbol{q}_n,\qquad
\boldsymbol{q}_n=\mathbf{U}\boldsymbol{a}_n.
\end{equation}

According to Eq.~\ref{equ:disdynamic}, projecting the dynamics onto the POD coordinates gives the evolution operator $g_\theta$ for the induced reduced dynamics
\begin{equation}
\boldsymbol{a}_{n+1}=g_\theta(\boldsymbol{a}_n)
=\mathbf{U}^*f_\theta(\mathbf{U}\boldsymbol{a}_n).
\end{equation}

Let $\boldsymbol{a}_b=\mathbf{U}^*\boldsymbol{q}_b$ be the base state expressed in the POD subspace. The Jacobian of the reduced dynamics $\mathbf{N}_\mathrm{POD}$ evaluated at $\boldsymbol{a}_b$ is
\begin{equation}
\mathbf{N}_{\mathrm{POD}}
=\frac{\partial g_\theta}{\partial \boldsymbol{a}}\Big|_{\boldsymbol{a}_b}
=\mathbf{U}^*\frac{\partial f_\theta}{\partial \boldsymbol{q}}\Big|_{\boldsymbol{q}_b}\mathbf{U}=\mathbf{U}^*\mathbf{N}\mathbf{U},
\end{equation}
where $\mathbf{N}$ is the original NN-based Jacobian. Since $\mathbf{U}$ defines an exact coordinate transform on the POD subspace, $\mathbf{N}_{\mathrm{POD}}$ is the representation of $\mathbf{N}$ in the POD coordinates restricted to this subspace.

Now let $\mathbf{V}_{\mathbf{N},s}=[\boldsymbol{v}_1,\cdots,\boldsymbol{v}_s]$ denote the dominant eigenvectors associated with the leading $s$ eigenvalues collected in $\boldsymbol{\Lambda}_{\mathbf{N},s}$, such that
\begin{equation}
\mathbf{N}\mathbf{V}_{\mathbf{N},s}=\mathbf{V}_{\mathbf{N},s}\boldsymbol{\Lambda}_{\mathbf{N},s}.
\end{equation}
If the POD subspace contains this leading invariant subspace,
$\mathrm{span}(\mathbf{V}_{\mathbf{N},s}) \subseteq \mathrm{range}(\mathbf{U})$,
then there exists a full-column-rank matrix $\mathbf{T}\in\mathbb{C}^{l\times s}$ such that
\begin{equation}
\mathbf{V}_{\mathbf{N},s}=\mathbf{U}\mathbf{T}.
\end{equation}
Substituting this into the eigenvalue relation and left-multiplying by $\mathbf{U}^*$ yields
\begin{equation}
\mathbf{N}_{\mathrm{POD}}\mathbf{T}
=\mathbf{T}\boldsymbol{\Lambda}_{\mathbf{N},s}.
\end{equation}

Therefore, the reduced dynamics expressed in the full POD coordinates preserve the eigenvalues of the original dynamics on the POD subspace, and the corresponding eigenvectors are related through the POD projection and lifting.

In practice, retaining a suitably chosen number of POD modes is usually adequate to recover the leading eigenvalues accurately, with the required number depending on the dimensionality of the specific problem. The reason is that POD ranks the modes according to their contribution to the snapshot energy, while the dominant linear structures are typically the most weakly damped or most amplified components and therefore dominate the variance of the trajectory data. As a result, these dynamically most important structures are preferentially captured by the leading POD modes, so that mode truncation usually has little effect on the dominant linear characteristics of interest. From a mathematical standpoint, the POD‑reduced system provides a Ritz approximation of the eigenspectrum of the original high‑dimensional system.  Hence, learning the truncated POD coefficient dynamics is sufficient to preserve the dominant linear stability characteristics while substantially reducing the cost of training and analysis.

\section{Results}\label{sec:applications}

We present four examples that highlight different features of the NN‑based modal analysis: a mean-field model of cylinder flow, the complex Ginzburg-Landau equation, a three-dimensional channel flow, and a real-world cyclostationary climate system.

To ensure a robust Jacobian estimation, we adopt a model ensemble strategy \citep{chua2018deep,allen2022inverse} by averaging the Jacobians predicted from five trained models initialized with different random seeds. This can not only mitigate the effects of weight initialization diversity but also lead to a more representative average gradient. The network architecture settings for each case, plotted in \textit{SI Appendix} Fig.~S15, were selected following the systematic experiments conducted in our prior benchmark study \citep{koehler2024apebench}. In addition, several ablation studies on data quality, as well as additional application examples, are also provided in \textit{SI Appendix} due to space constraints.

\subsection{Mean-field model of cylinder flow}

\begin{sidewaystable}
\centering
\begin{tabular}{ccccc}
\hline \addlinespace[2pt]
                          & \multicolumn{2}{c}{analytical results} & \multicolumn{2}{c}{NN-based results} \\ \hline \addlinespace[2pt]
linear stability analysis & Jacobian $\exp(\mathbf{A}\Delta t)$     & eigenvalues    & Jacobian $\mathbf{N}$       & eigenvalues       \\ \addlinespace[2pt] \hline \addlinespace[5pt]
fixed point $\begin{bmatrix} 0 \\ 0 \\ 0 \end{bmatrix}$               & $\begin{bmatrix} 1.0038&-0.0502&0 \\ 0.0502&1.0038&0 \\ 0&0&0.6065\end{bmatrix}$ & $\begin{bmatrix} 0.6065 \\ 1.0038 \pm 0.0502i\end{bmatrix}$ & $\begin{bmatrix} 1.0038&-0.0502&-0.0001\\0.0502&1.0038&0\\-0.0002&-0.0002&0.6059\end{bmatrix}$ & $\begin{bmatrix}0.6059\\1.0038 \pm 0.0502i\end{bmatrix}$ \\  \addlinespace[5pt]
non-fixed point $\begin{bmatrix} -0.4697\\-0.2437\\0.1023 \end{bmatrix}$          & $\begin{bmatrix} 1.0028&-0.0504&0.0018 \\ 0.0499&1.0031&0.0010 \\ -0.3755&-0.1821&0.6060 \end{bmatrix}$ & $\begin{bmatrix} 0.6082\\ 1.0019 \pm 0.0503i \end{bmatrix}$ & $\begin{bmatrix} 1.0022&-0.0506&0.0012\\ 0.0493&1.0024&-0.0002\\ -0.3715&-0.1923&0.6051\end{bmatrix}$ & $\begin{bmatrix} 0.6061\\1.0018 \pm 0.0504i \end{bmatrix}$ \\  \addlinespace[5pt] \hline \addlinespace[2pt]
Floquet analysis          & Monodromy $M$           & eigenvalues    & Monodromy $\prod_{n=m}^1 \mathbf{N}_n$      & eigenvalues       \\ \addlinespace[2pt] \hline \addlinespace[5pt]
\makecell{periodic orbit \\ $z=x^2+y^2=1$}             & $\begin{bmatrix} 0.2832&0&-0.0029\\ 0&1&-0.0289\\  0.5781&0&-0.0059\end{bmatrix}$ & $\begin{bmatrix}1\\0.2772\\0\end{bmatrix}$ & $\begin{bmatrix} 0.2813&-0.0168&-0.0027\\ 0.0027&0.9999&0.0002\\ 0.5744&0.0002&-0.0055
\end{bmatrix}$ & $\begin{bmatrix}0.9998\\ 0.2758\\ 0\end{bmatrix}$ \\ \addlinespace[5pt] \hline
\end{tabular}
\caption{Comparison of NN‑based and operator-based results for linear stability and Floquet analyses. The linear stability analysis is performed at the fixed point and a representative non‑fixed point (the initial condition shown in Fig.~\ref{fig:meanmodel}), while the Floquet analysis is performed on the periodic orbit with the same starting point.}
\label{tab:meanfield}
\end{sidewaystable}

We first consider a system that possesses both a fixed point and a periodic orbit, so that linear stability analysis and Floquet analysis are applicable. We adopt the mean-field model of the laminar cylinder wake \citep{noack2003hierarchy,lusch2018deep}, a three-state Galerkin system reproducing the transient growth of von K\'arm\'an vortex shedding and its nonlinear saturation onto a periodic orbit. The state $\boldsymbol{q}=[x,y,z]^T$ is governed by

\begin{equation}\label{equ:meanfield}
\begin{aligned}
\dot{x} &= \mu x - \omega y + \gamma x z, \\
\dot{y} &= \omega x + \mu y + \gamma y z, \\
\dot{z} &= -\lambda\left(z - x^2 - y^2\right),
\end{aligned}
\end{equation}

\noindent where $(x,y)$ are the oscillatory shedding amplitudes and $z$ the mean-field distortion, slaved for large $\lambda$ to the slow manifold $z = x^2+y^2$. With the standard parameters $\mu = 0.1$, $\omega = 1$, $\gamma = -0.1$, $\lambda = 10$, the origin is an unstable fixed point, while the saturated dynamics collapses onto a stable periodic orbit $\boldsymbol{q}_b(t) = [\cos t, \sin t, 1]^T$ of period $T = 2\pi$. 

By applying a local linearization to Eq.~\ref{equ:meanfield} around a base state $\boldsymbol{q}_b = [x_b, y_b, z_b]^T$ , we obtain the analytical local Jacobian:

\begin{equation}\label{equ:Jmeanfield}
\mathbf{A} = \frac{\partial \mathcal{N}}{\partial \boldsymbol{q}}\Big|_{\boldsymbol{q}_b} =
\begin{bmatrix}
\mu+\gamma z_b & -\omega & \gamma x_b \\
\omega & \mu+\gamma z_b & \gamma y_b \\
2\lambda x_b & 2\lambda y_b & -\lambda
\end{bmatrix},
\end{equation}

For a fixed‑point base state, this directly gives the ground‑truth Jacobian, whereas for a limit‑cycle base state, the ground‑truth monodromy matrix is obtained via time‑exponential integration based on Eq.~\ref{equ:monodromy_product}.

\begin{figure}[t]
    \centering
    \includegraphics[width=\linewidth]{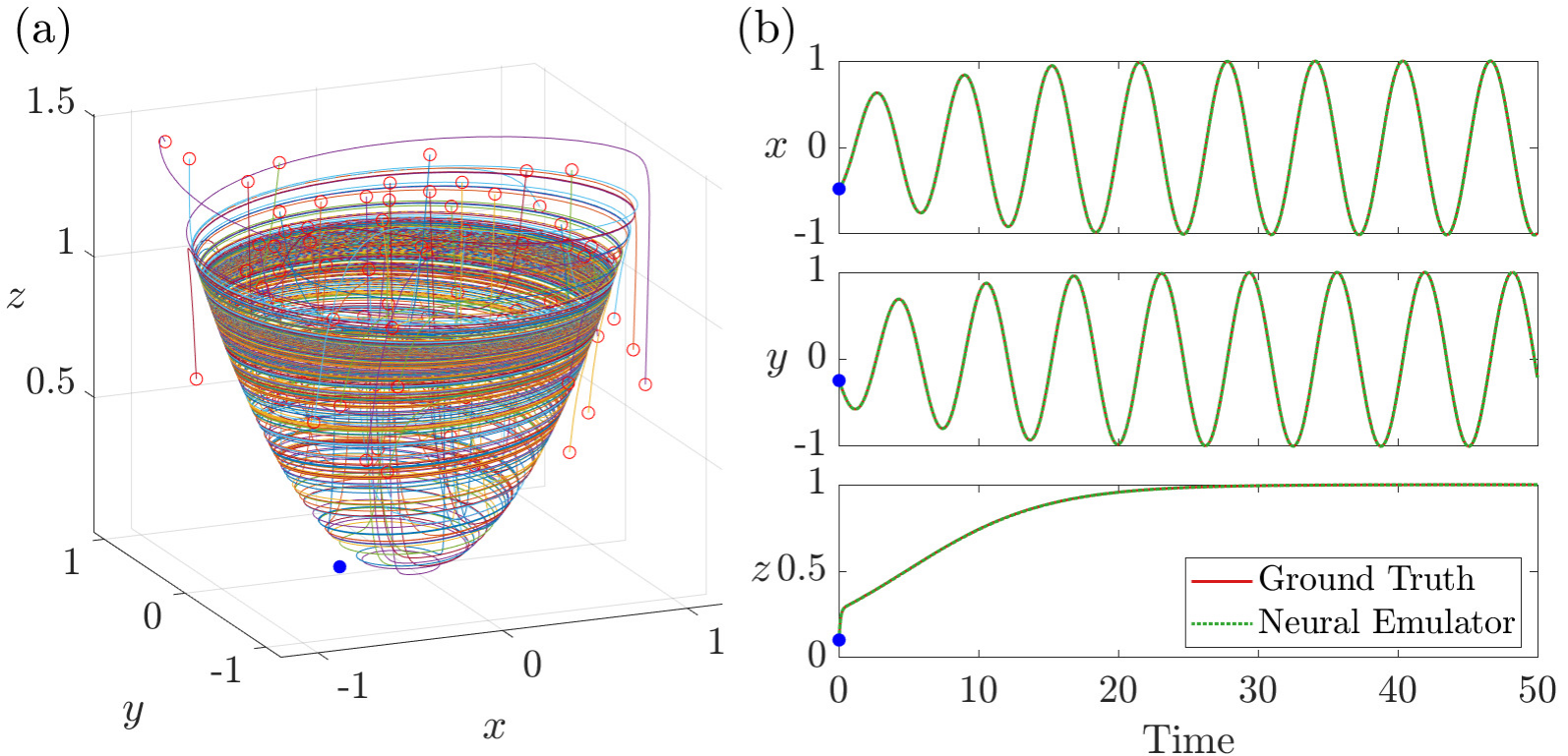}
    \caption{(a) 100 trajectories of the mean‑field model from randomly initialized points (red circles). The blue dot marks  an unseen initial point used for testing. (b) Component‑wise comparison of the ground‑truth trajectory and the NN emulator prediction from the unseen initial point.}
    \label{fig:meanmodel}
\end{figure}

The training data in Fig.~\ref{fig:meanmodel}(a) consist of 100 trajectories initialized randomly and integrated for 1000 steps using a fourth‑order Runge–Kutta scheme with a fixed time step $\Delta t = 0.05$. Although this step size is not commensurate with the period $T=2\pi$, it is sufficiently small to yield a good approximation of the monodromy matrix. The NN architecture is a three-layer network with 10 neurons per hidden layer.

Despite being trained on the trajectory dataset with only a one-step rollout strategy, the network can be used to autoregressively generate new trajectories as far into the future as desired, even for initial conditions not present in the training dataset. Fig.~\ref{fig:meanmodel}(b) presents the component-wise trajectory comparison between the NN emulator's prediction and ground truth starting from a randomly selected initial condition over an interval of 50 time units.  The close agreement between both simulations in the early stage suggests that the trained NN emulator captures the system dynamics with high fidelity, effectively mimicking the update rule of a classical Runge–Kutta scheme over a time increment of $0.05$. 

Meanwhile, we leverage the automatic differentiation to perform linear stability analysis and Floquet analyses for the fixed point and periodic orbit, and compare them with the analytical results. As shown in Tab.~\ref{tab:meanfield}, the NN-based and analytical Jacobian and monodromy matrices exhibit an almost perfect element-wise match, and consequently their eigenvalues are nearly identical. Although the incommensurability between $\Delta t$ and $T$ introduces a minor discretization error, it remains negligible for the small $\Delta t$ employed here.

At the origin, the Jacobian possesses an eigenvalue with magnitude greater than one, indicating that the fixed point is unstable. For the periodic orbit, by contrast, the Floquet multipliers all lie inside the unit circle, indicating that the periodic orbit is orbitally stable. Although the NN emulator is trained purely on data and is agnostic to the underlying governing equations, these results suggest that it can not only accurately capture the evolution of the system (zeroth-order information), but also faithfully reproduce the perturbation dynamics (derivative-level information) in the vicinity of both the fixed point and the periodic orbit. This paves the way for data-driven modal analysis approaches for both linear and nonlinear scenarios in higher-dimensional settings.

Furthermore, we demonstrate that for this simple ODE system, the Jacobian agreement is not limited to the fixed point but instead extends to arbitrary points along the trajectory (second row of Tab.~\ref{tab:meanfield}). This indicates that the trained neural operator can also faithfully represent local dynamical behavior at any points and may facilitate tasks such as computing finite-time Lyapunov exponents (FTLE) \citep{shadden2005definition}. This is not the primary focus of our investigation.

\subsection{Complex Ginzburg-Landau system}

The second example considered is the one-dimensional complex Ginzburg–Landau equation \citep{bagheri2009input}. It is widely recognized as a reduced-order model for studying diverse instability behaviors in spatially evolving flows by varying parameter configurations. As its linearized operator can be obtained analytically through the manipulation of the differential matrices, it has become a benchmark for the evaluation and development of stability analysis techniques \citep{chen2011h2,herrmann2021data,martini2021efficient}.

The nonlinear complex Ginzburg–Landau equation defined on an infinite interval $x \in (-\infty,\infty)$ is written as

\begin{equation}
    \frac{\partial \boldsymbol{q}}{\partial t} = \left(-\nu \frac{\partial}{\partial x}+\mathbf{\gamma}\frac{\partial^2}{\partial x^2}+\mu(x)\right)\boldsymbol{q}-a|\boldsymbol{q}|^2\boldsymbol{q} \\
    = \mathbf{A}\boldsymbol{q}-a|\boldsymbol{q}|^2\boldsymbol{q},
\end{equation}

\noindent where $-a|\boldsymbol{q}|^2\boldsymbol{q}$ is the cubic nonlinear term. Since $\boldsymbol{q}=0$ is an equilibrium point, the linearized system around this point simply involves omitting the cubic term and thus $\mathbf{A}$ is the linearized operator. 

The advection and dispersion properties are controlled by the complex terms $\nu =U + 2ic_u$ and $\gamma = 1 + ic_d$, respectively. The real-valued term $\mu(x)=\mu_0-c_u^2+\mu_2 x^2/2$ is defined as a quadratic function to model exponential instabilities \citep{bagheri2009input}. 
The homogeneous boundary conditions are considered at $x \to \pm\infty$, so that perturbations are allowed to grow and decay throughout the entire domain. 

\begin{table}
  \centering
  \begin{tabular}{c@{\hskip 30pt}l@{\hskip 30pt}c}
      Variable  & Description  &   Value \\[3pt] \midrule
      $U$ & Mean advection velocity & 2.0 \\
      $c_u$ & Most unstable wavenumber & 0.2 \\
      $c_d$ & Dispersion parameter & -1.0 \\
      $\mu_0$ & Bifurcation parameter & 0.23(0.41) \\
      $\mu_2$ & Degree of non-parallelism & -0.01 \\
      $a$ & Nonlinearity & 0(1) \\ \bottomrule
  \end{tabular}
  \caption{Parameter descriptions and settings for the complex Ginzburg–Landau system. Values corresponding to the nonlinear scenario are indicated in parentheses.}
  \label{tab:gl}
\end{table}

\begin{figure}
  \centerline{\includegraphics[width=\linewidth]{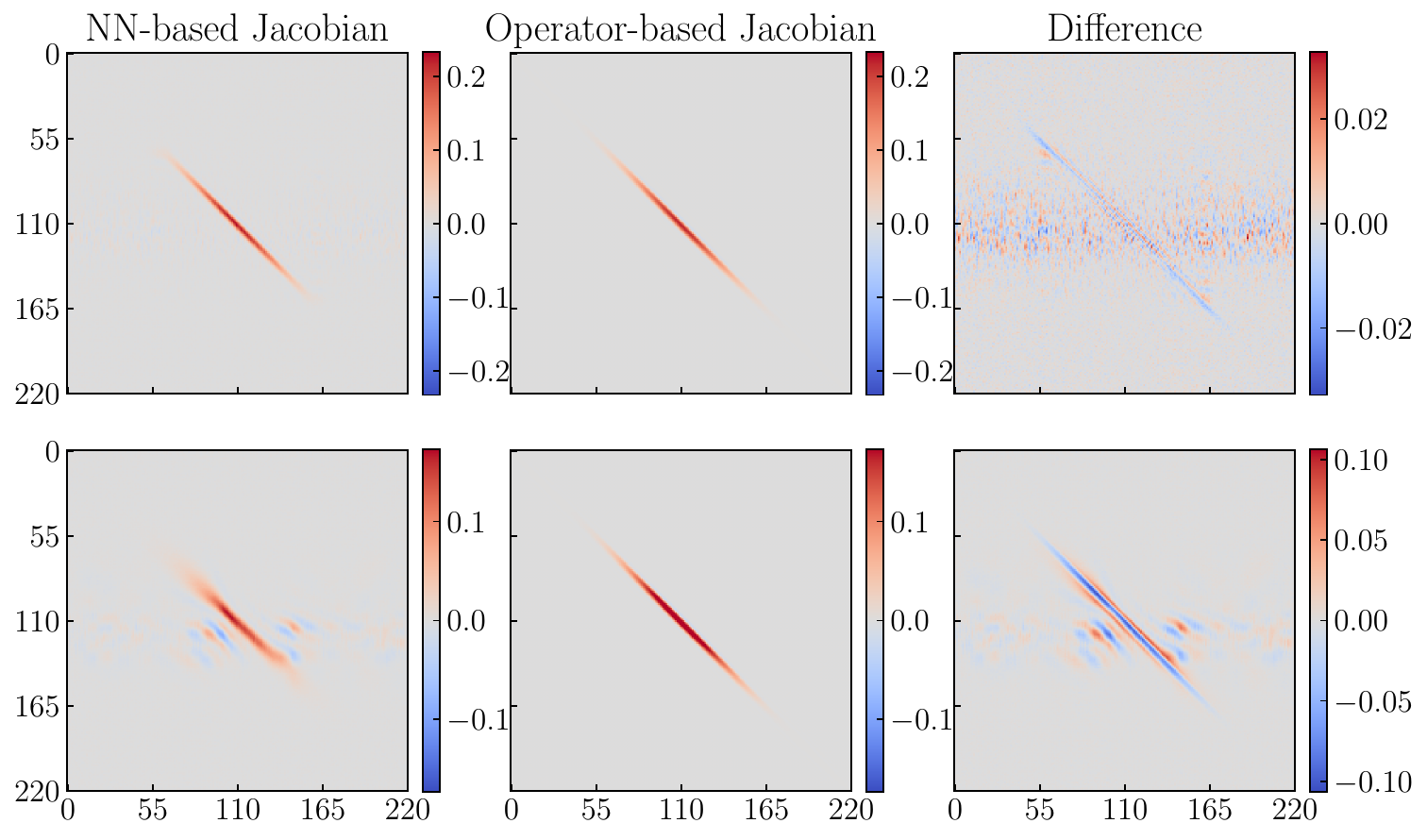}}
  \caption{Comparison of the real part of the Jacobians between the NN-based operator and the operator-based ground truth. The first and second rows correspond to the linear and nonlinear scenarios, respectively.}
  \label{fig:jacobian}
\end{figure}

We aim to demonstrate the applicability of the NN emulator approach to both linear and nonlinear regimes even in complex number space. For this purpose, we consider two datasets derived respectively from the linear \citep{herrmann2021data} and nonlinear \citep{bagheri2009input} settings of this system. The sample trajectories and the specific parameter settings with their descriptions are presented in \textit{SI Appendix}, Fig.~S1 and Tab.~\ref{tab:gl}.

For both scenarios, we conduct 30 independent simulations, each initiated with distinct initial conditions derived from the superposition of multiple random Gaussian profiles. For each simulation, $m=100$ snapshots are collected at regular time intervals of $0.5$. We employ a complex-valued, six-layer perceptron with $512$ neurons per hidden layer. A truncation level of $r=24$ is used for the eigenbasis subspace in this example. All results are obtained using a one-step rollout training strategy. More details about this case are provided in \textit{SI Appendix}, Sec.~2.

As shown in Fig.~\ref{fig:jacobian}, we compare the Jacobian heatmaps of the trained emulator with those of the analytical operator for both linear and nonlinear scenarios. We can find that the network perfectly recovers the operator in the linear scenario, whereas for the nonlinear scenario, it successfully captures the main features but exhibits minor discrepancies in finer details. These deviations are most likely caused by nonlinear mappings being inherently more difficult to approximate than linear ones.

\begin{figure}
  \centerline{\includegraphics[width=\linewidth]{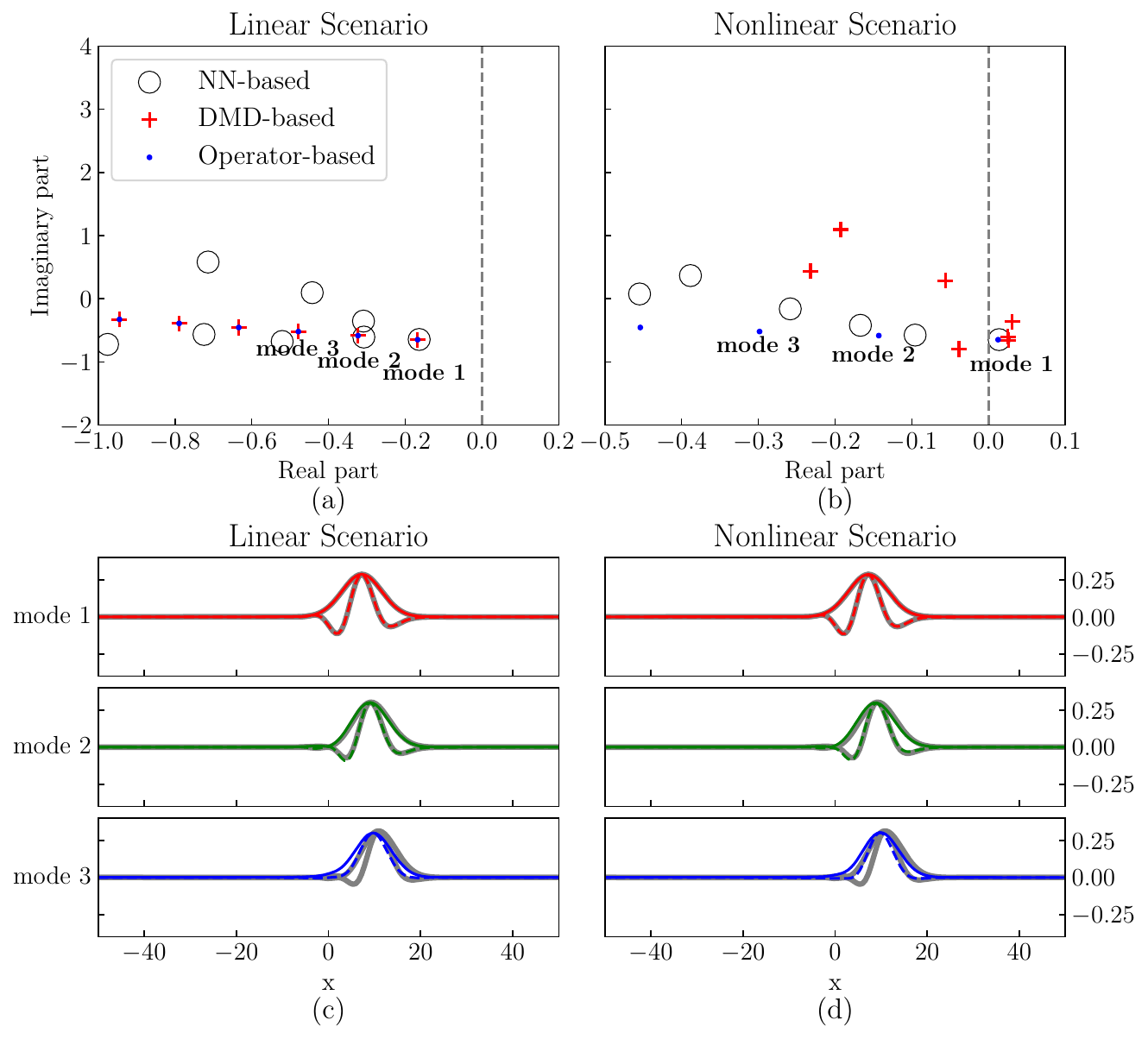}}
  \caption{Linear stability analysis results of the linear and nonlinear Ginzburg-Landau systems. (a)(b) Eigenspectra obtained from the NN-based Jacobian ($\circ$), DMD-based Jacobian ({\color{red}$+$}) and operator-based ground truth ($\color{blue}\bullet$). The dashed line denotes the stability boundary in the complex plane. (c)(d) The leading three eigenmodes of the NN-based Jacobians in both scenarios, where solid and dashed lines show the real part and magnitude of the modes. The thick gray lines in the background show operator-based ground-truth for comparison.}
  \label{fig:cgle_eigenspectrum}
\end{figure}

In the context of linear stability analysis, Fig.~\ref{fig:cgle_eigenspectrum}(a) and (b) compare the eigenspectra of the NN-based and operator-based Jacobians for linear and nonlinear scenarios, respectively. For reference, we also include eigenspectra from the DMD model, computed using the exact DMD approach \citep{tu2014dynamic}. Fig.~\ref{fig:cgle_eigenspectrum}(c) and (d) show the leading three eigenmodes identified by the NN emulator for both scenarios. For the linear scenario, it is anticipated that the DMD-based results show excellent agreement with the analytical solution, as DMD is fundamentally designed to approximate linear operators. Even so, the NN-based results also show excellent alignment of the leading eigenvalues and eigenvectors with the analytical reference. For the nonlinear scenario, DMD-based approach breaks down entirely, as demonstrated by the identification of more than one unstable mode located in the right half of the complex plane (see Fig.~\ref{fig:cgle_eigenspectrum}(b)). In contrast, the NN-based Jacobian still yields an accurate leading unstable eigenvalue, even though it differs in finer structures according to Fig.~\ref{fig:jacobian}. This suggests that the NN emulator can effectively identify the most dominant mode regardless of the dynamical properties of datasets. While the higher-order modes exhibit stronger noise‑related fluctuations, they decay rapidly and contribute minimally to the long-term behavior of perturbations.

\begin{figure}[t!]
  \centerline{\includegraphics[width=\linewidth]{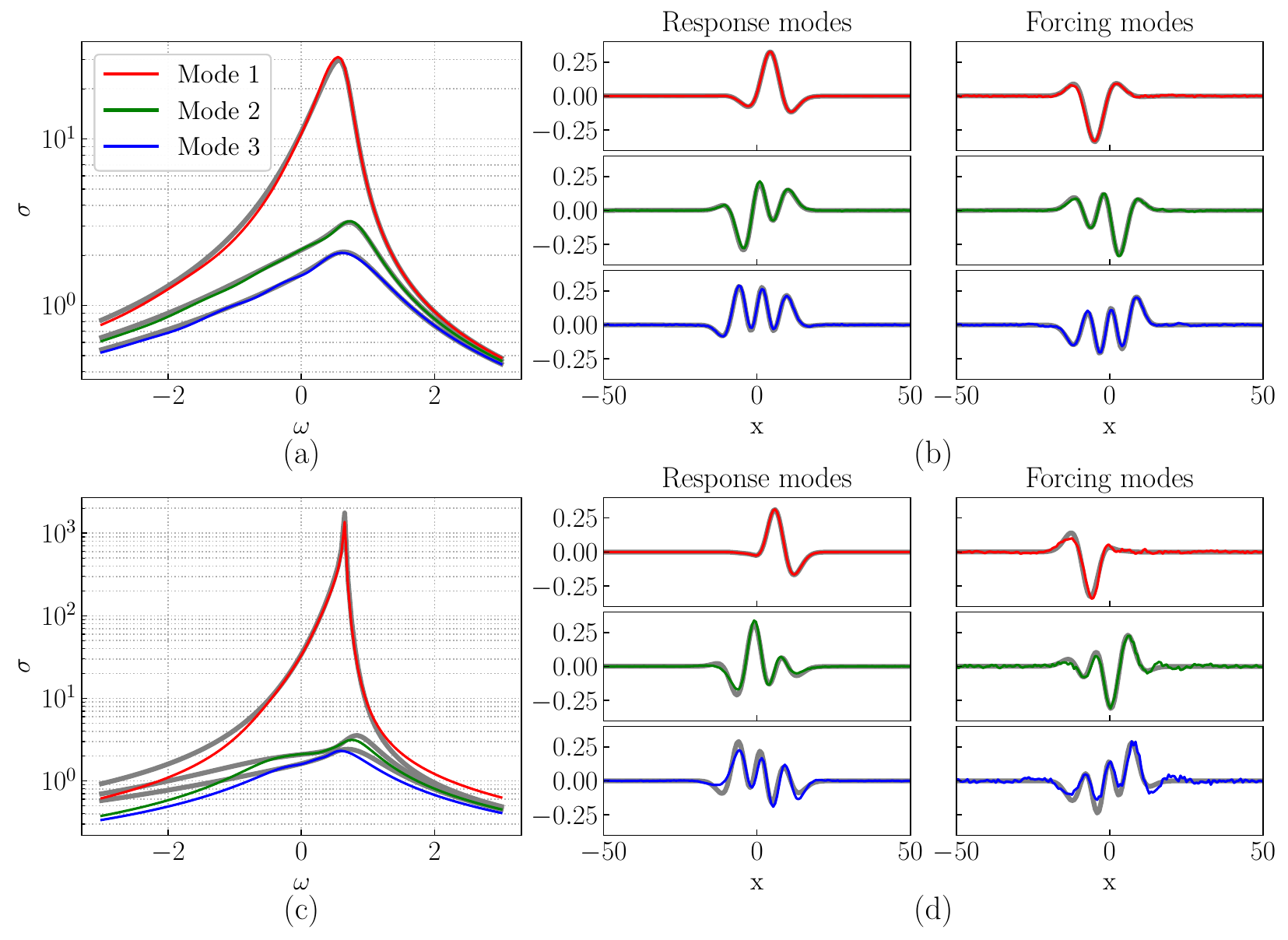}}
  \caption{Resolvent analysis results of the linear and nonlinear complex Ginzburg–Landau systems. (a) Resolvent gain distribution of the first three modes as a function of frequency for the linear system. (b) First three forcing and response modes at the peak-gain frequency for the linear system. (c and d) As in A and B but for the nonlinear system. Thick gray lines in the background show the operator-based ground truth for comparison.}
  \label{fig:cgle_resolvent}
\end{figure}

In the context of resolvent analysis, Fig.~\ref{fig:cgle_resolvent}(a) and (c) show the gain curves of the first three resolvent modes as a function of the forcing frequency for the two scenarios. The corresponding dominant response and forcing modes at the peak gain frequency are presented in Fig.~\ref{fig:cgle_resolvent}(b) and (d). For the linear scenario, our NN-based approach can yield highly accurate gain curves and mode shapes. For the nonlinear scenario, it also succeeds in predicting the gain curve near the dominant frequency, and accurately reconstructs the response and forcing modes corresponding to this frequency. 
Though applicable to both linear and nonlinear regimes, the higher-order mode shapes predicted by the NN emulator become increasingly influenced by noise and deviate from ground truth more noticeably, as shown in Fig.~\ref{fig:cgle_resolvent}(c) and (d).

There are two important aspects to note for the gain distributions of Fig.~\ref{fig:cgle_resolvent}(c).
First, the NN emulators have worse performance for gain prediction at frequencies far from the peak gain. This phenomenon can be interpreted via dyad expansion of the resolvent operator \citep{symon2018non}. The resolvent operator $H(\omega)$ can be expressed as a sum of dyadic products of left eigenvectors $\tilde{\boldsymbol{g}}_{j}$ and right eigenvectors $\tilde{\boldsymbol{h}}_{j}$, each weighted by $1/(-i \omega - \lambda_j)$ as

\begin{equation}\label{equ:rank1}
\mathbf H(\omega)=(-i\omega\mathbf{I}-\mathbf{A})^{-1}=\sum_{j=1}^{n}\frac{1}{-i\omega-\lambda_{j}}\tilde{\boldsymbol{g}}_{j}\tilde{\boldsymbol{h}}_{j}^{*}\approx \frac{1}{-i\omega-\lambda_{1}}\tilde{\boldsymbol{g}}_{1}\tilde{\boldsymbol{h}}_{1}^{*},
\end{equation}

\noindent where we can see that near the frequency of maximal gain, i.e., the forcing frequency $\omega$ is close to the imaginary part of the dominant eigenvalue, the resolvent operator admits a good rank-1 approximation. Since the leading eigenmode is accurately recovered (see Fig.~\ref{fig:cgle_eigenspectrum}), the corresponding resolvent operator is also highly accurate for those frequency intervals satisfying the rank-1 approximation.

Second, the NN-based method tends to exhibit more noise in the predicted forcing modes compared to the response modes. However, this behavior is in line with other data-driven approaches and can be attributed to the same cause: in non-normal systems, the spatial structure of the direct eigenvectors resembles that of the response modes, but can differ significantly for the forcing modes \citep{herrmann2021data}.

\subsection{Reduced 3D transitional channel flow}

To further demonstrate the applicability of our method to high‑dimensional systems, we consider the three‑dimensional channel flow  with Re = 2000 (based on the centerline velocity and the channel half-height) governed by the incompressible Navier-Stokes equations. By considering perturbations in the wavenumber space, the analytical derivation of the linearized operator can be effectively reduced to the Orr–Sommerfeld (OS) equation \citep{schmid2012stability}, which we use for the ground truth reference (Details in \textit{SI Appendix}, Sec.~3). 

In this case, the NN emulator aims to model the temporal evolution of infinitesimal perturbations around the base parabolic velocity profile, and hence, snapshots of the velocity perturbation fields are collected during the transient evolution stage for training. 
The dataset is generated by a spectral solver \citep{shenfun,mortensen_joss} on a computational domain of $2\pi \times 2 \times 2$ in the streamwise, spanwise and wall-normal directions, where periodicity is imposed along the streamwise and spanwise direction. The spatial domain is discretized by $N_y = 64$ Chebyshev grids in the wall-normal direction and $N_x=32$ Fourier grids in the streamwise and spanwise direction. 

To satisfy the incompressibility constraint, the initial velocity perturbation fields in the three-dimensional domain are prescribed as

\begin{equation}
\begin{aligned}
u(x,y,z) &= \epsilon \cdot \Re\left[\phi(y)e^{\mathrm{i}(k_x x + k_z z)}\right],\\
v(x,y,z) &= \epsilon \cdot \Re\left[\frac{\mathrm{i}k_x\,\phi'(y)-\mathrm{i}k_z\,\phi(y)}{k_x^2+k_z^2}e^{\mathrm{i}(k_x x + k_z z)}\right], \\
w(x,y,z) &= \epsilon \cdot \Re\left[\frac{\mathrm{i}k_z\,\phi'(y)+\mathrm{i}k_x\,\phi(y)}{k_x^2+k_z^2}e^{\mathrm{i}(k_x x + k_z z)}\right].
\end{aligned}
\end{equation}

The initial profile $\phi(y)$ is constructed as a superposition of multiple Chebyshev polynomial profiles with random amplitudes, with special treatment at the wall boundaries to satisfy both homogeneous Dirichlet and Neumann boundary conditions. The streamwise and spanwise wavenumbers are fixed as $k_x=k_z=1$ for simplicity. And $\epsilon$ scales the initial perturbation and thus determines the strength of nonlinear interactions, which has a pronounced impact on the transient flow behavior. We consider a strongly nonlinear dataset with $\epsilon = 0.1$. A total of $60$ trajectories are simulated with $m=400$ snapshots saved every $0.5$ time units.

Although a convolutional NN emulator can in principle learn the snapshot‑to‑snapshot mapping in physical space, the resulting Jacobian matrix has a dimension equal to the total number of degrees of freedom of all flow variables over the full discretized field. For this case, the three-dimensional field corresponds to a state vector of dimension close to $2\times 10^5$, which makes direct eigenvalue decomposition and the subsequent construction of the resolvent operator computationally expensive. Hence, reduced-order models are attractive to make such problems tractable. 

The POD reduction is applied to the stored data snapshots to obtain the first 100 orthogonal spatial modes and corresponding temporal coefficient trajectories. Then, we train the NN emulator on these coefficient trajectories and apply the following modal analysis in an encoder-decoder approach. Since the temporal coefficients form a one-dimensional vector, we employ a four-layer multilayer perceptron with 256 neurons in each hidden layer. The truncation level used for the projection eigensubspace in the resolvent analysis is chosen to be identical to the number of POD modes, and thirty-step temporal unrolling is used during training.

\begin{figure}[t]
  \centerline{\includegraphics[width=\linewidth]{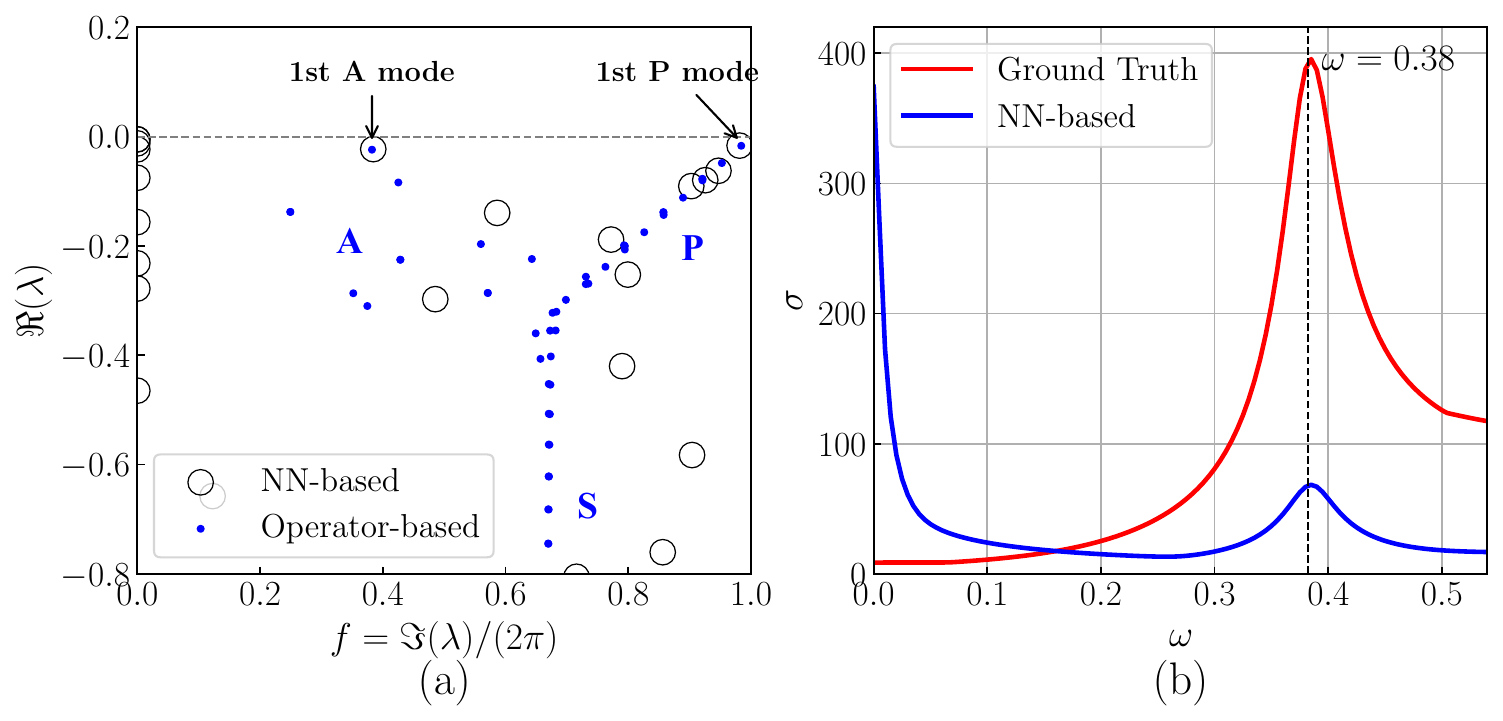}}
  \caption{Comparison of the (a) eigenspectra and (b) resolvent gain curves from the operator-based Jacobian and NN-based Jacobian for the 3D channel flow system.}
  \label{fig:chan3d}
\end{figure}

For the chosen Reynolds number of 2000, the channel system is stable, which can be verified by the eigenspectrum of the OS operator from Fig.~\ref{fig:chan3d}(a). All operator-based eigenvalues fall in the stable complex half-plane and they are located on three branches forming a Y-shaped pattern. 
Based on the phase speed $c_r$, these branches have been labeled as A ($c_r \rightarrow 0$), P ($c_r \rightarrow 1$), and S ($c_r \approx 2/3$) by \citep{mack1976numerical}. The branches describe distinct features of the perturbation dynamics: P modes (center modes) on the P-branch with much higher phase speeds capture the fast dynamics in the center of the channel, and A modes (wall modes) with rather small phase velocities capture the slow dynamics near the channel wall. The S modes approaching a phase speed $2/3$ are highly damped.

\begin{figure}[t]
  \centerline{\includegraphics[width=\linewidth]{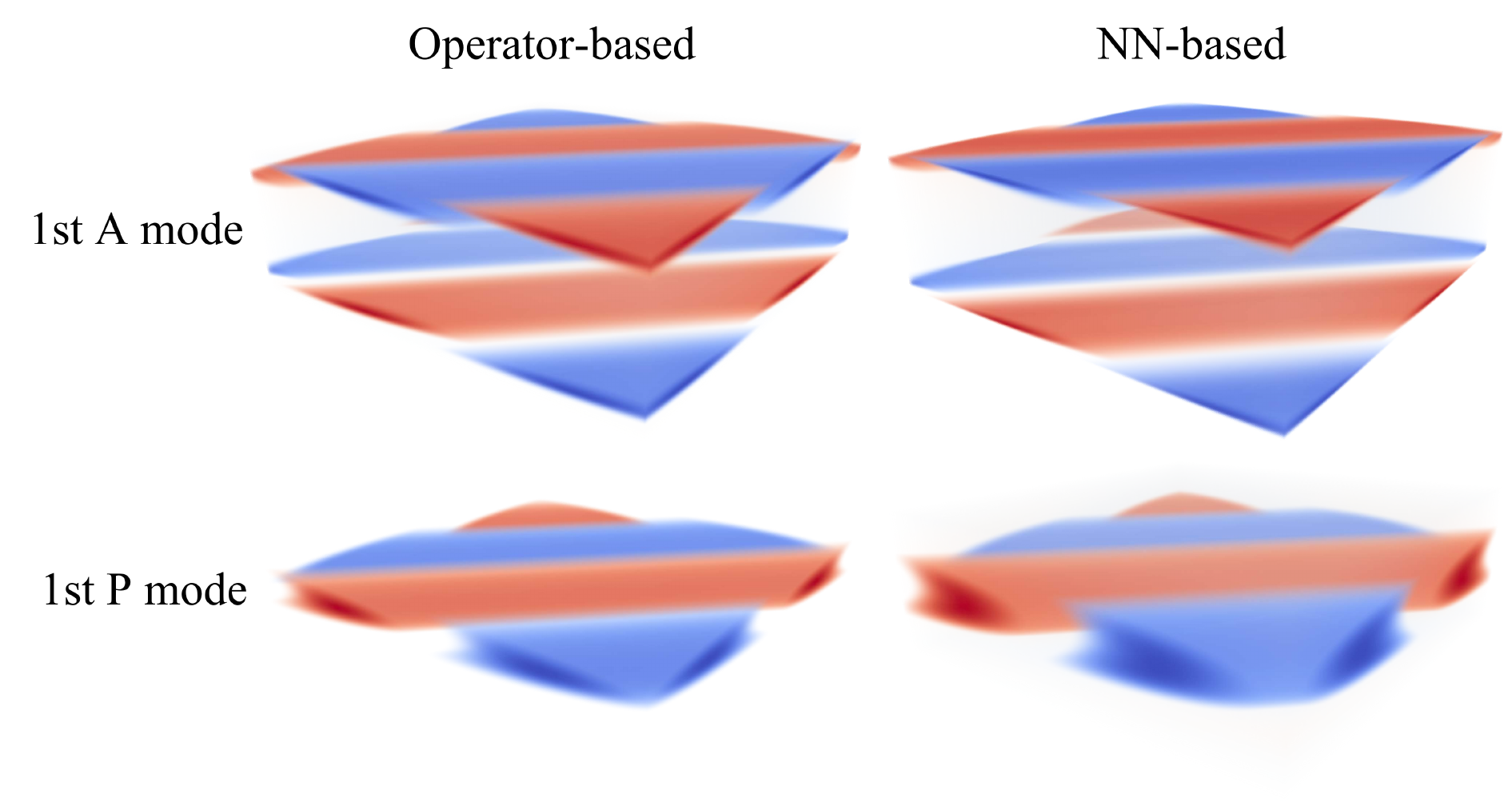}}
  \caption{Comparison of the first-order A and P eigenmodes of the streamwise velocity field for the 3D channel flow system.}
  \label{fig:chan3d_eigenmode}
\end{figure}

In Fig.~\ref{fig:chan3d}(a), we note that the leading eigenvalues of the A-branch and P-branch agree well between the operator-based and NN-based results. Despite the strong nonlinearities of the dataset, the NN emulator can successfully identify the two most dominant perturbation structures near the wall and centerline, i.e., the least-damped A and P modes. Their spatial modes presented in Fig.~\ref{fig:chan3d_eigenmode} appear as oblique waves which are a set of inclined wave-like structures traversing the flow field \citep{jovanovic2005componentwise}. Such oblique waves are known to play an important role because they can experience transient amplification through the coupling between wall-normal velocity and vorticity perturbations via a vortex-stretching mechanism.

The emulator, though trained on reduced POD coefficients, still captures the leading linear stability characteristics of the original full-state system. While the S modes with high decay rates were not recovered, their contribution to the system's dynamics is negligible. We attribute this to limitations in the numerical precision of the dataset, as well as the inherent noise in the trained emulator, which potentially obscures the evolution of these highly damped modes.

\begin{figure}[t]
  \centerline{\includegraphics[width=\linewidth]{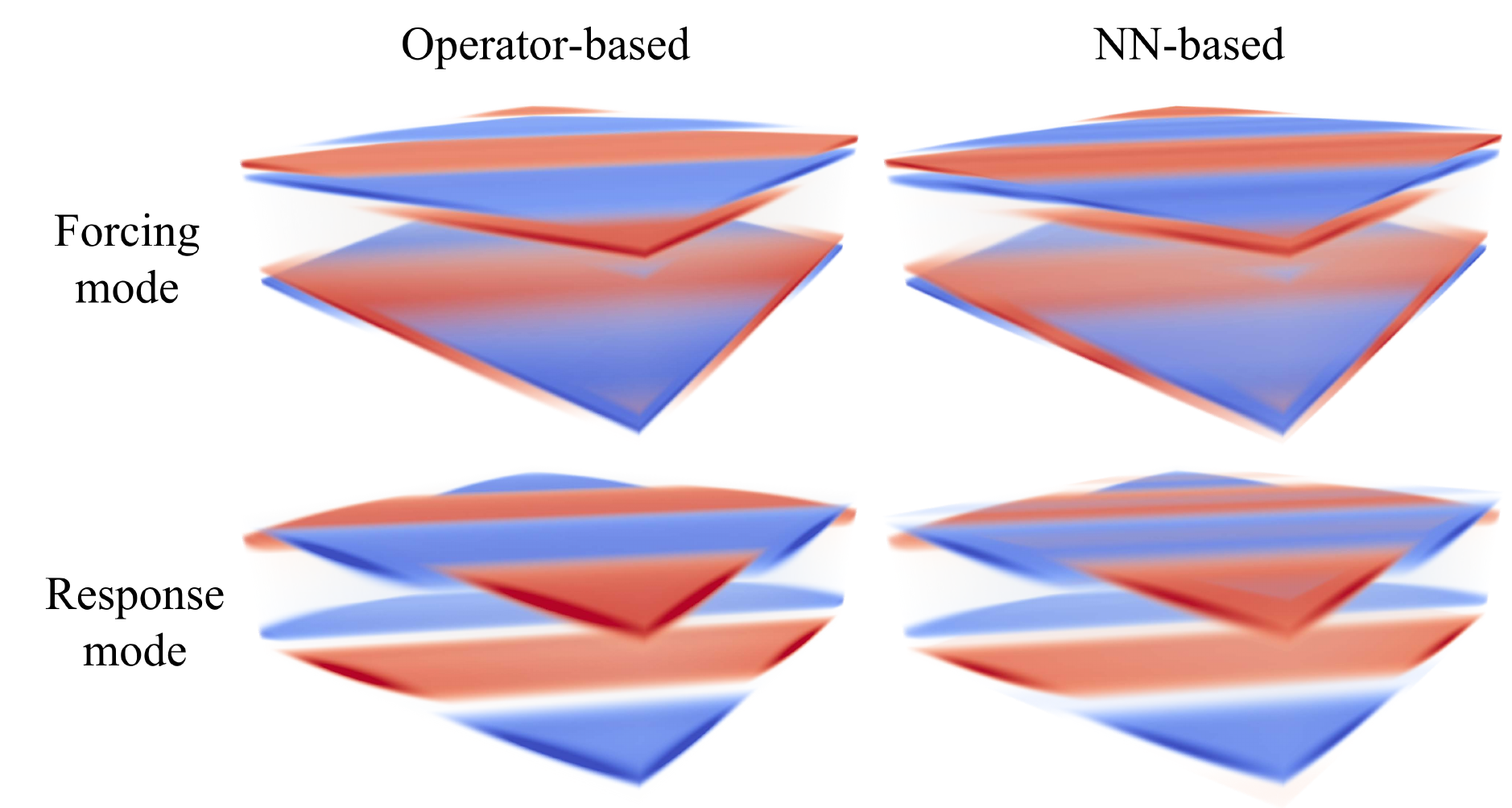}}
  \caption{Comparison of the first-order forcing and response modes of the streamwise velocity field at $\omega = 0.38$ for the reduced 3D channel flow system. }
  \label{fig:chan3d_resolvent}
\end{figure}

Meanwhile, several secondary A and P modes fall along the branch of the analytical results, which is not the case when trained directly on the full-state dataset. A contrasting behavior is observed for the 2D channel flow case (\textit{SI Appendix}, Fig.~S8), where training on the full state can capture only the leading A and P modes and shows no discernible branch structure. Although secondary modes correspond to more rapidly decaying dynamics, POD separates modes into different energy levels beforehand. Rather than forcing the neural emulator to discover these less pronounced structures directly from the full state, learning the dynamics within the POD modal subspace is apparently more efficient.

It is noteworthy that many spurious eigenvalues appear on the imaginary axis (Fig.~\ref{fig:chan3d}(a)). This discrepancy is attributed to the preprocessing applied to the training data. The raw data were preprocessed by subtracting the base flow and then normalizing the result. While such treatment is reasonable in the linear or weakly nonlinear regime, it can introduce distortions in strongly nonlinear scenarios, where the principle of linear superposition no longer holds. This effect is also confirmed in the 2D channel flow case, in which we generated highly nonlinear and weakly nonlinear datasets by varying the coefficient $\epsilon$. As a consequence, zero-frequency modes associated with non-oscillatory components are introduced into the data and are subsequently captured by the NN emulator, manifesting as eigenvalues along the imaginary axis in Fig.~\ref{fig:chan3d}(a).

As for the resolvent analysis, Fig.~\ref{fig:chan3d}(b) displays the gain curves of two distinct operators in response to harmonic forcings with frequencies in the range of $[0, 0.5]$. 
The peak frequency at $\omega = 0.38$ is accurately predicted by the NN emulator despite the nonlinear nature of this scenario.
We note that the corresponding resolvent gain curve exhibits an extra peak at zero frequency for the nonlinear scenario. This phenomenon aligns with the presence of non-oscillatory spurious modes discussed above, and the additional gain peak comes from resonating with them according to Eq.~\ref{equ:rank1}. In our tests, it can be substantially mitigated either by filtering the corresponding spurious modes in the reconstructed spectrum or by training on the wall-normal velocity dataset alone, for which no such subtraction is required. Nevertheless, the NN emulator accurately predicts the physically relevant peak frequency of $\omega = 0.38$ originating from a near-resonance with the first A-branch eigenmode.

The resolvent modes of streamwise velocity perturbations at $\omega = 0.38$ are illustrated in Fig.~\ref{fig:chan3d_resolvent}, revealing that the NN emulator detects the area that is most sensitive to perturbation amplification for channel flow at the current Reynolds number. Since the leading eigenvectors of the NN-based Jacobian align closely with those of the true system, it naturally leads to accurate resolvent mode predictions. Similarly, the reconstructed response mode shows near-exact agreement with the ground truth, whereas the forcing mode yields a less accurate approximation, which is also attributed to the non-normal nature of the channel flow. 

In addition to reducing the dimensionality of the learning problem, POD reduction offers several further advantages. Traditional convolutional  architectures can only enlarge the receptive field gradually through stacked encoder blocks, while POD modes already provide global spatial patterns of the dominant flow structures. The neural network thus only needs to learn the dynamics within the subspace spanned by these physically relevant modes, which is particularly important for fluid systems with global spatial interactions. Moreover, the extracted POD modes for channel flow usually inherit the symmetry or central symmetry with respect to the channel centerline. Therefore, the reconstructed stability and resolvent modes, expressed as combinations of POD modes, provide a powerful inductive bias to preserve the expected symmetry.

\subsection{Cyclostationary climate system}

We finally consider a high-dimensional, equation-free climate dataset without ground-truth governing equations, specifically sea surface temperature (SST) and sea surface height (SSH). Its latent dynamics  behave like those of a seasonally varying system whose periodic base state is the seasonal cycle itself, $\boldsymbol q_b(t)=\boldsymbol q_b(t+T)$ with $T=12$ months, making it a natural high-dimensional test of the Floquet analysis. Because an analytical Jacobian is unavailable, we adopt as a reference from the climate community the traditional data-driven linear inverse model (LIM) \citep{penland1995optimal} and its cyclostationary version (CSLIM) \citep{vimont2022role}.

A distinguishing feature of this case is the presence of stochastic noise in the dataset. LIM usually models the SST/SSH anomaly (perturbation) $\boldsymbol q'$ as a linear stochastic differential equation,

\begin{equation}\label{equ:sde}
\mathrm d\boldsymbol q' = \mathbf A\,\boldsymbol q'\,\mathrm dt + \mathrm d\boldsymbol\xi(t), \quad \langle\mathrm d\boldsymbol\xi\rangle = 0,
\end{equation}

\noindent where $\boldsymbol\xi$ is the stochastic term. The dynamical matrix $\mathbf A$ is constant for LIM and $T$-periodic for CSLIM with $\mathbf A(t+T)=\mathbf A(t)$. CSLIM estimates the monthly propagators via $\mathbf G_n=\mathbf C_n(\Delta t)\,\mathbf C_n(0)^{-1}$, where $\mathbf C_n(\tau)=\mathbb{E}[\boldsymbol{q}'_{n+\tau}\boldsymbol{q}'^T_n]$ represents the lag‑$\tau$ covariance matrix at month $n$. Therefore, the noisy component only contributes to the covariances and cancels through $\mathbf C_n(0)^{-1}$. 

Our NN emulator, by contrast, learns the conditional mean of SST/SSH full-state transition from month $n$ to $n+1$ via

\begin{equation}\label{equ:condmean}
f_\theta = \arg\min_{f_\theta}
\mathbb{E}\bigl\|\mathbf{q}_{n+1} - f_\theta(\mathbf{q}_n)\bigr\|^2
\Longrightarrow
f_\theta(\mathbf{q}_n) \approx \mathbb{E}[\mathbf{q}_{n+1}\,|\,\mathbf{q}_n],
\end{equation}

\noindent so that noise in the training targets is averaged out by the L2-norm loss, and it enters neither the Jacobian nor the monodromy. Both routes therefore share the same deterministic, $T$-periodic operator $\mathbf A(t)$, and their results of Floquet analysis are directly comparable.

To allow for direct comparisons, we use an existing dataset \citep{vimont2022role}: 
SST from the HadISST dataset \citep{rayner2003global} and SSH from the ECMWF Ocean Reanalysis System  \citep{balmaseda2013evaluation} over the region $100^\circ$E–$75^\circ$W, $25^\circ$S–$25^\circ$N over the period 1958–2017. Following the common practice in meteorology, the two-dimensional latitude-longitude fields are also reduced by POD (9 leading principal components for SST and 5 leading principal components for SSH), and the temporal coefficients are normalized before training. 

In contrast to the previous examples,  the climate dataset spans only sixty years, providing 720 snapshots in total. To avoid overfitting under such limited sampling, we adopt a simple two‑layer residual network with 64 units per layer and train an ensemble of ten networks. A long rollout training strategy is not advisable for the this case, as rollout training would tend to force the NN emulator to match the entire noisy trajectory and compel it to explain the noise, which yields a poor Jacobian estimate. We therefore train with a one-step rollout.

\begin{figure}
    \centering
    \includegraphics[width=\linewidth]{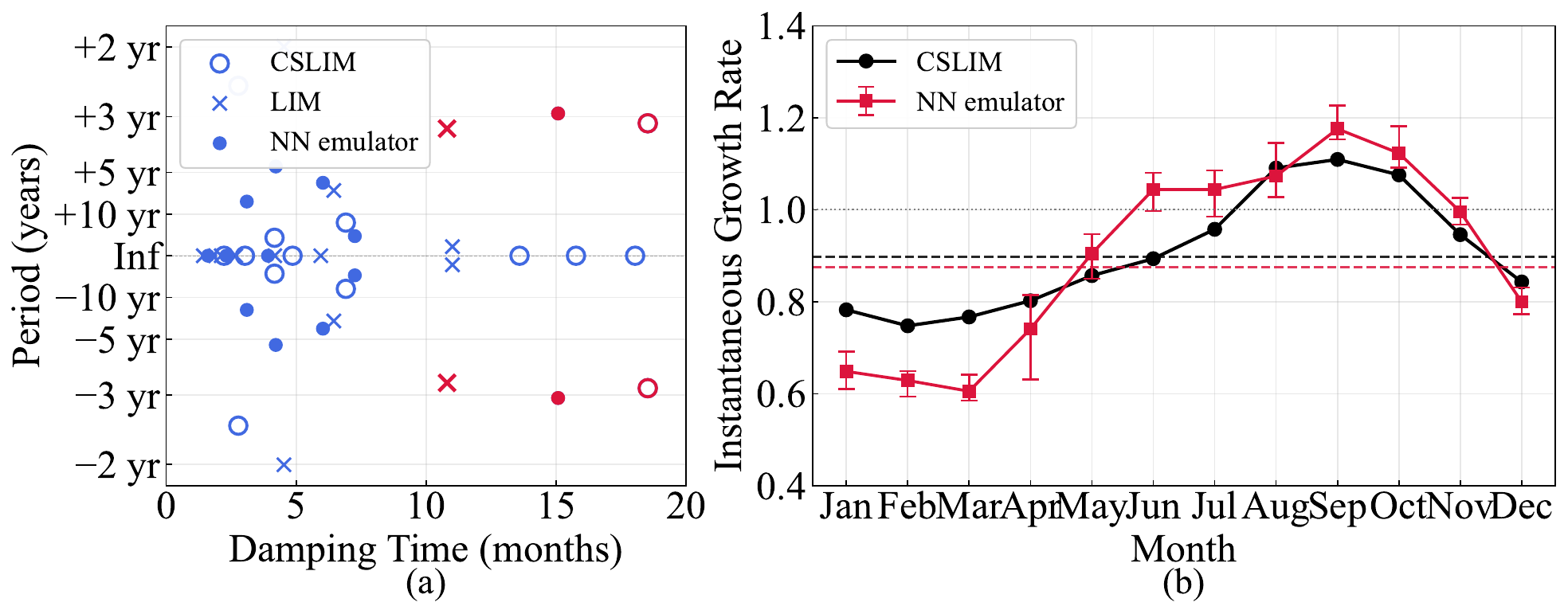}
    \caption{(a) Floquet spectrum from CSLIM (circles) and NN emulator (dots) and eigenspectrum (crosses) from LIM. The red symbols indicate the leading ENSO mode predicted by each method. (b) Monthly varying instantaneous growth rates of the seasonally varying ENSO mode from CSLIM and NN emulator, as well as the geometric mean of these growth rates (horizontal dashed line). }
    \label{fig:enso_spec}
\end{figure}

\begin{figure}
    \centering
    \includegraphics[width=\linewidth]{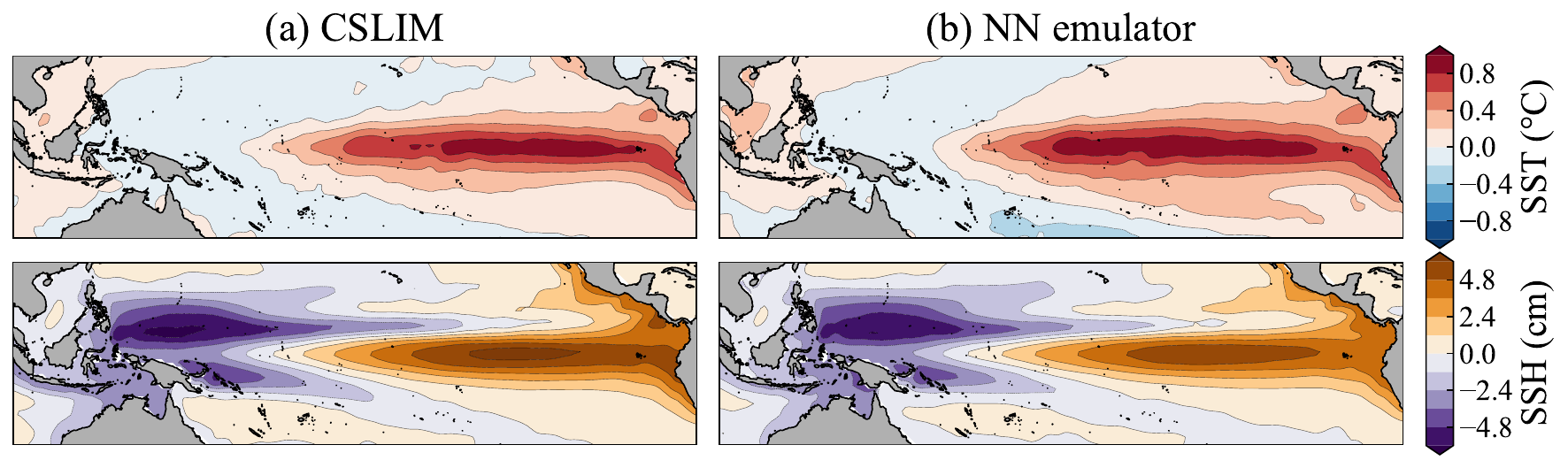}
    \caption{Spatial structure of the real component of the ENSO mode when evaluated along the annual cycle starting from December, where the first row denotes SST and the second row denotes SSH.}
    \label{fig:enso_mode}
\end{figure}

The Floquet spectrum from CSLIM and NN emulator are shown in Fig.~\ref{fig:enso_spec}(a) together with the eigenspectrum of the stationary LIM. All contain a conjugate pair of oscillatory modes with nearly identical oscillation periods of around 3 years and very similar damping time scales (11 months for LIM, 19 months for CSLIM, and 15 months for the NN emulator). Given their agreement with known El Ni\~no-Southern Oscillation (ENSO) time characteristics, they are identified as the ENSO mode. Fig.~\ref{fig:enso_spec}(b) also shows the instantaneous growth rate, which is quantified via one-month propagation of the normalized ENSO mode. The seasonality of the growth rates is well captured by the NN emulator. The seasonal periodic state enhances the thermocline–SST coupling during summer, producing a transient growth rate exceeding one, which overcomes the ENSO mode's net annual damping. Despite local overestimates or underestimates of growth rates for a few months, the overall growth rate of the ENSO mode (dashed horizontal line) is essentially unbiased, with values of 0.90 for CSLIM and 0.88 for LIM. 

The real components of ENSO modes for December are shown in Fig.~\ref{fig:enso_mode}. Our predicted modes are likewise in close agreement with CSLIM results for both SST and SSH data, exhibiting deepened equatorial thermocline anomalies that extend across the Pacific and evolve eastward into a surface ENSO event. Additional results and comparison are given in \textit{SI Appendix}, Sec.~4. The emulator thus recovers the ENSO-mode dynamics directly from observations, demonstrating that the present data‑driven framework generalizes beyond low‑dimensional toy models to high‑dimensional, noise‑contaminated geophysical systems, with promising potential for broader future applications.

\section{Discussion and Outlook}\label{sec:conclusion}

In this work, we have developed a data-driven framework for extracting dominant local linear characteristics from a differentiable NN emulator. Rather than requiring explicit access to the governing equations or a tractable linearized operator, the method learns the nonlinear flow map directly from trajectory data and then recovers local Jacobian-based information through automatic differentiation. In this way, stability analysis and resolvent analysis can be extended to equation-free settings, where conventional operator-based approaches are difficult or unavailable.
Its effectiveness is demonstrated on four representative systems of increasing complexity and nonlinearity, underscoring its potential for analyzing complex dynamics. The fact that such results are achieved purely through the representational capacity of primitive network architectures is highly encouraging. 

We also compared our framework with the typical data-driven approach, DMD. The results show a clear regime dependence. When the training data are generated by linear or weakly nonlinear dynamics, DMD yields an accurate eigenspectrum reconstruction, which is expected because it is fundamentally designed to approximate linear evolution operators directly from data. As the data move further into a fully nonlinear regime, DMD loses even the dominant spectral structure, whereas our framework can still retain sufficient local dynamical information to recover the leading Jacobian-based characteristics. Although an exact one-to-one reconstruction of the full eigenspectrum is challenging, the dominant local linear characteristics are usually most important and physically relevant.

Despite these encouraging results, several limitations also delineate the current scope. First, the data must cover the neighborhood of the base state, and for noisy observations the optimal rollout horizon is problem dependent rather than monotonically increasing. Second, our reduced-order formulation is accurate only within the retained subspace, so the basis must preserve dynamically relevant directions. Third, all current experiments assume full-state observations, while under partial observability, the learned Jacobian reflects only the projected dynamics, whose spectrum may not match that of the underlying system. Recovering the full physical modes generally requires additional treatments.

As for future improvements, previous extensions and modifications to the DMD approach may offer valuable insights for enhancing our approach. One promising direction lies in incorporating sparsity‑promoting constraints to automatically select the few most dynamically relevant modes, in a manner analogous to sparsity‑promoting DMD\citep{jovanovic2014sparsity}. Moreover, since DMD can handle non-uniformly sampled datasets to circumvent the Nyquist criterion \citep{gueniat2015dynamic}, our NN emulator could also be extended to such data by adopting higher-order integration schemes rather than the direct mapping used in this work.
 
On the other hand, NN-based modal analysis also provides a perspective for assessing the fidelity of a trained NN emulator in capturing the system's true dynamics or for diagnosing prediction divergence through Jacobian‑based stability analysis, especially when reference modal results are available. This synergy between machine learning and classical modal analysis opens a promising avenue toward interpretable, physics-informed modeling and advances the integration of data-driven and theoretical approaches.

\appendix
\renewcommand{\thefigure}{S\arabic{figure}}
\renewcommand{\thetable}{S\arabic{table}}
\setcounter{figure}{0}
\setcounter{table}{0}
\renewcommand{\theequation}{S\arabic{equation}}
\setcounter{equation}{0}
\section*{Supporting Information}

\section{Modal analysis}
\subsection{Linear stability analysis}

When there is no exogenous forcing applied to the original dynamical system, the governing equation of the perturbation $\boldsymbol{q}'$ about a base state $\boldsymbol{q}_b$ degrades to a homogeneous system:

\begin{equation}\label{equ:dynamic_nof}
    \frac{d\boldsymbol{q}'}{dt}=\mathbf{A}\boldsymbol{q}',
\end{equation}

\noindent where we can assume the perturbation in the form of normal mode $\boldsymbol{q}'=\hat{\boldsymbol{q}}e^{\sigma t}$ with $\sigma=\lambda+i \omega$. The real part $\lambda$ and the imaginary part $\omega$ are the growth rate and characteristic frequency, respectively, and $\hat{\boldsymbol{q}}$ describes the mode shape. By substituting the normal mode into Eq.~\ref{equ:dynamic_nof}, we finally obtain an eigenvalue problem as follows:

\begin{equation}
\mathbf{A}\hat{\boldsymbol{q}}=\sigma \hat{\boldsymbol{q}}.
\end{equation}

Since the Jacobian matrix $\mathbf{A}$ depends on the base state $\boldsymbol{q}_b$, different points along the trajectory generally exhibit different local stability properties, which depends on whether each corresponding $\mathbf{A}$ has at least one eigenvalue with a positive real part $\lambda$. 

\subsection{Floquet analysis}
The linear stability analysis assumes the base state $\boldsymbol{q}_b$ to be a certain point. However, many physical systems instead settle onto a time-periodic orbit. In this case, the base state satisfies $\boldsymbol{q}_b(t)=\boldsymbol{q}_b(t+T)$ with period $T$, and the stability of such an orbit is characterized by Floquet analysis, which can be regarded as the extension of linear stability analysis from fixed points to periodic orbits.

Linearizing the dynamics about the periodic orbit $\boldsymbol{q}_b(t)$, the perturbation $\boldsymbol{q}'$ is governed by

\begin{equation}\label{equ:dynamic_periodic}
    \frac{d\boldsymbol{q}'}{dt}=\mathbf{A}(t)\boldsymbol{q}',\qquad \mathbf{A}(t)=\mathbf{A}(t+T),
\end{equation}

\noindent where the Jacobian $\mathbf{A}(t)$ is now intrinsically $T$-periodic. Eq.~\ref{equ:dynamic_periodic} no longer admits normal-mode solutions of a single exponential form. Instead, its solution is governed by the monodromy matrix $\mathbf{M}$, which maps a perturbation forward by exactly one period,

\begin{equation}\label{equ:monodromy}
    \boldsymbol{q}'(t+T)=\mathbf{M}\,\boldsymbol{q}'(t),\qquad \mathbf{M}=\mathcal{T}\exp\!\int_{t}^{t+T}\mathbf{A}(\tau)\,d\tau, \qquad
    \mathbf{M}\hat{\boldsymbol{q}}=\lambda \hat{\boldsymbol{q}},
\end{equation}

\noindent where $\mathcal{T}\exp$ denotes the time-ordered exponential. Floquet analysis then consists of the eigendecomposition of $\mathbf M$: the Floquet modes are $\hat{\boldsymbol{q}}$, the Floquet multipliers are $\lambda_k = \mathrm{eig}(\mathbf M)$ and the Floquet exponents are $\mu_k = T^{-1}\log\lambda_k$. By construction, there is one trivial multiplier equal to unity, with its eigenvector tangent to the orbit. The orbit is asymptotically stable when all remaining multipliers lie inside the unit circle $|\lambda| < 1$, and becomes unstable once any of them crosses it.

\subsection{Resolvent analysis}
Stability analysis provides only a partial description of the system dynamics because it solely focuses on unstable modes. In practice, a system can remain asymptotically stable with all eigenvalues residing in the stable region, yet still exhibit significant transient growth of perturbations. This motivates the use of tools such as resolvent analysis to obtain an input-output viewpoint of the perturbation evolution. 

Since we aim to examine the response outputs of perturbation $\boldsymbol{q}'$ for different forcing inputs and identify the optimal input mode which yields the most amplified energy by internal dynamics, the external forcing term $\boldsymbol{f}$ should be taken into account. Thus, the evolution of the perturbation $\boldsymbol{q}'$ is described by a forced linear dynamical system as:

\begin{equation}\label{equ:dynamic_f}
    \frac{d\boldsymbol{q}'}{dt}=\mathbf{A}\boldsymbol{q}'+\boldsymbol{f}.
\end{equation}

We assume the response of perturbations $\boldsymbol{q}'$ and the external forcing $\boldsymbol{f}$ to be harmonic, which can be expressed as $\boldsymbol{q}'(t) = \hat{\boldsymbol{q}}e^{-i\omega t}$ and $\boldsymbol{f}(t) = \hat{\boldsymbol{f}}e^{-i\omega t}$, where $\omega$ is the angular driving frequency. Based on this assumption, Eq.~\ref{equ:dynamic_f} is rewritten as 

\begin{equation}\label{equ:resolvent}
\hat{\boldsymbol{q}}=(-i\omega \mathbf{I}- \mathbf{A})^{-1}\hat{\boldsymbol{f}}=\mathbf{H}(\omega) \hat{\boldsymbol{f}},
\end{equation}

\noindent where $\mathbf{H}(\omega) \in \mathbb{C}^{N \times N}$ is defined as the resolvent operator, i.e., transfer function matrix between the forcing inputs and response outputs at frequency $\omega$. 

Since the goal is to obtain the optimal energy amplification over all possible forcing modes $\hat{\boldsymbol{q}}$, a physically meaningful norm for energy measurement needs to be selected. A state variable norm can be defined as $|| \hat{\boldsymbol{q}} ||^2_{\mathbf{Q}}=\hat{\boldsymbol{q}}^*\mathbf{Q}\hat{\boldsymbol{q}}$, where $()^*$ denotes the Hermitian transpose and $\mathbf{Q}=\mathbf{F}^*\mathbf{F}$ is a positive-definite weighting matrix that may account for grid discretization size \citep{yuan2023resolvent}, compressible flow energy \citep{chu1965energy} and other factors. The $\mathbf{Q}$ norm is related to the standard Euclidean norm as $|| \hat{\boldsymbol{q}} ||^2_{\mathbf{Q}}=|| \mathbf{F}\hat{\boldsymbol{q}} ||^2_2$. The largest input-output gain can be quantified via the ratio of their respective norms

\begin{equation}\label{equ:gain}
\sigma^2(\omega)=\max_{\hat{\boldsymbol{f}}} \frac{|| \hat{\boldsymbol{q}'} ||^2_{\mathbf{Q}}}{|| \hat{\boldsymbol{f}} ||^2_{\mathbf{Q}}} = \max_{\hat{\boldsymbol{f}}} \frac{|| \mathbf{H}(\omega) \hat{\boldsymbol{f}} ||^2_{\mathbf{Q}}}{|| \hat{\boldsymbol{f}} ||^2_{\mathbf{Q}}} = ||\mathbf{F}\mathbf{H}(\omega)\mathbf{F}^{-1}||^2_2.
\end{equation}

Eq.~\ref{equ:gain} is an optimization problem, whose solution is given by the weighted SVD of the resolvent operator

\begin{equation}
\mathbf{F}\mathbf{H}(\omega)\mathbf{F}^{-1} = \boldsymbol{\Psi}_{\mathbf{F}}(\omega) \boldsymbol{\Sigma}(\omega) \boldsymbol{\Phi}^*_{\mathbf{F}}(\omega),
\end{equation}

\noindent where $\boldsymbol{\Sigma} \in \mathbb{R}^n$ is the singular value matrix whose diagonal entries are the optimal gains associated with the leading $n$ resolvent modes, and the column vectors in $\mathbf{F}^{-1}\boldsymbol{\Psi}_{\mathbf{F}}(\omega)$ and $\mathbf{F}^{-1}\boldsymbol{\Phi}_{\mathbf{F}}(\omega)$ are the corresponding response and forcing modes.

\section{Complex Ginzburg-Landau equation}
\subsection{Supercritical and subcritical behaviors}
\begin{figure}
\centering
\includegraphics[width=0.8\linewidth]{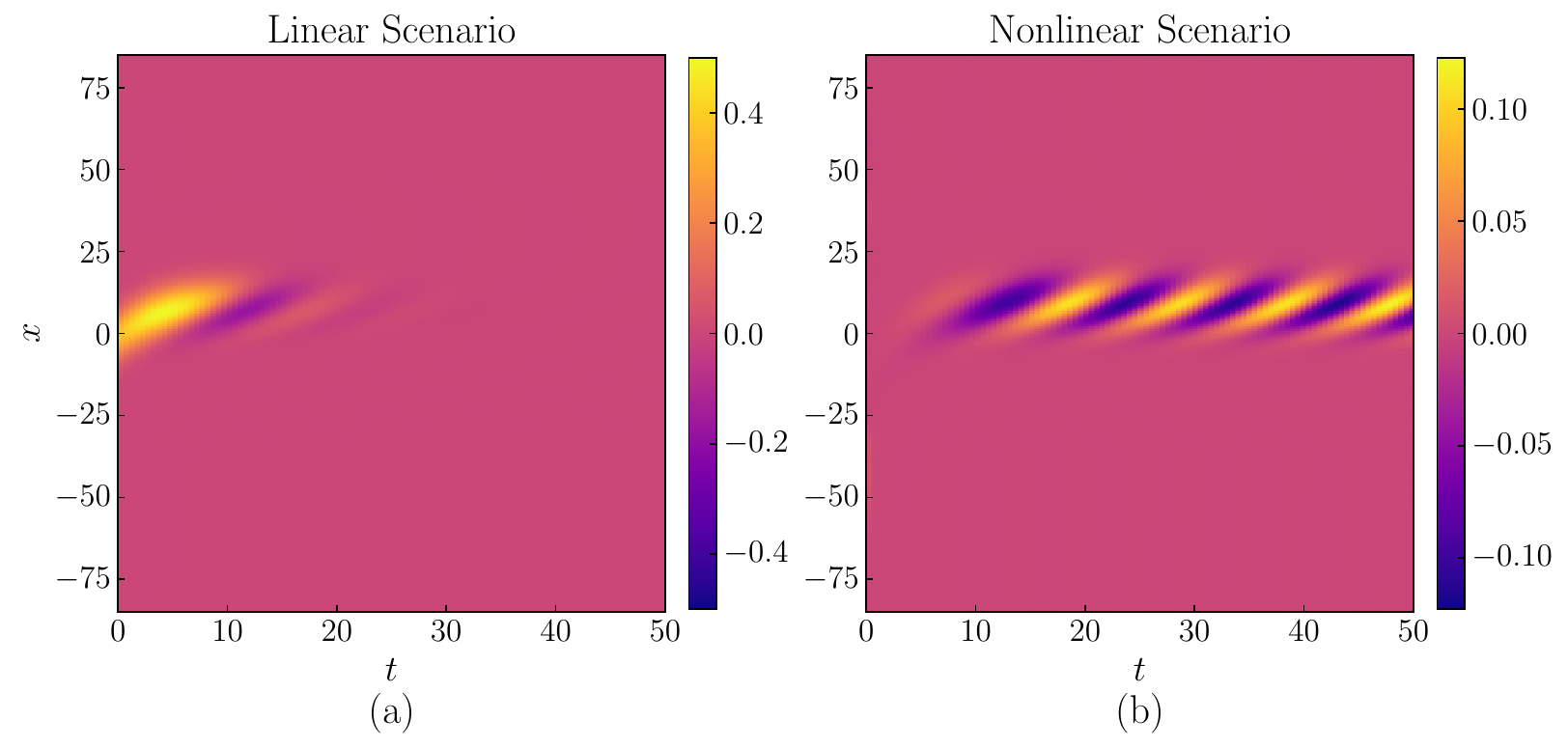}
\caption{Sample trajectories from datasets of (a) linear and (b) nonlinear complex Ginzburg-Landau systems.}\label{fig:cgle}
\end{figure}

The Ginzburg–Landau equation can describe two fundamentally different behaviors of real flows, depending on the comparison between the bifurcation parameter $\mu_0$ and the critical value for global instability $\mu_c=0.4$ for our parameter settings. 
The subcritical regime ($0<\mu_0<\mu_c$) is often referred to as a noise amplifier, in which the advection is so strong that all perturbations will be convected downstream before amplification. This corresponds to the convectively unstable flows, such as boundary layers \citep{ehrenstein2005two} and mixing layers \citep{ho1984perturbed}. 
The supercritical regime ($\mu_0>\mu_c$) corresponds to self-sustained oscillatory behavior, where perturbations will grow exponentially until the nonlinear cubic term causes them to saturate and oscillate, as shown in Fig.~\ref{fig:cgle}(b). This can be used to model global unstable flows, such as hot jets \citep{lesshafft2006nonlinear} and cylinder wakes \citep{lauga2004performance,cohen2005closed}.

\subsection{Discrete representation}
To discretize the linearized operator $\mathbf{A}$ of the Ginzburg-Landau equation, we first need to create a domain discretization using Hermite functions \citep{bagheri2009input}. A total of $N=220$ collocation points ${x_1, x_2, \cdots, x_N }$ are given by the roots of $H_n(\chi x)$, where $H_n$ is the \textit{n}-th Hermite polynomial and $\chi=(-\mu_2/(2\gamma))^{1/4}$ is a scaling factor. With this choice of N, the discretized domain stretches from $x_1 = -84.99$ to $x_N = 84.99$. Then the corresponding Hermite differentiation matrices provided by \citep{weideman2000matlab} are adopted to formulate the first derivative $\partial/\partial x$ and second derivative $\partial^2/\partial x^2$ to construct the discrete matrix $\mathbf{A}$ of size $N\times N$.

\section{Hydrodynamic stability}
\subsection{Linearized Navier-Stokes equations}
The motion of a three-dimensional incompressible flow is governed by the Navier-Stokes equations in nondimensional form (characteristic velocity \(U\) and length \(L\) are used to nondimensionalize) as:

\begin{equation}\label{equ:ns}
\frac{\partial \boldsymbol{u}}{\partial t} + \boldsymbol{u} \cdot \nabla \boldsymbol{u} = -\nabla p + Re^{-1} \nabla^2 \boldsymbol{u}, ~~~~ \nabla \cdot \boldsymbol{u} = 0.
\end{equation}

\noindent where $\boldsymbol{u}$ is the velocity field, $p$ is the pressure and $Re$ is the Reynolds number. In accordance with the conventional framework of linear stability analysis, we can linearize Eq.~\ref{equ:ns} around a steady base state $(\boldsymbol{U}_b, P_b)$ to obtain the linearized Navier-Stokes equations governing perturbations $(u', p')$:

\begin{equation}\label{equ:lin_ns1}
\frac{\partial \boldsymbol{u}'}{\partial t} + \boldsymbol{U}_b \cdot \nabla \boldsymbol{u}' + \nabla \boldsymbol{U}_b \cdot \boldsymbol{u}' = -\nabla p' + Re^{-1} \nabla^2 \boldsymbol{u}', ~~~~ \nabla \cdot \boldsymbol{u}' = 0.
\end{equation}

Based on aforementioned normal mode expansions $(\boldsymbol{u}'(x, y, t), p'(x, y, t))^T = (\hat{\boldsymbol{u}}(x, y), \hat{p}(x, y))^T e^{\sigma t}$ with $\sigma = \lambda + i\omega$, we substitute it into the \eqref{equ:lin_ns1}, which then can be reformulated in matrix form:

\begin{equation}
    \mathbf{A} \begin{pmatrix}
        \hat{\boldsymbol{u}}\\ \hat{p}
    \end{pmatrix}=\sigma\begin{pmatrix}
        \hat{\boldsymbol{u}}\\ \hat{p}
    \end{pmatrix},\quad\mathrm{with} ~~\mathbf{A}=
\begin{pmatrix}
-\boldsymbol{U}_b\cdot\nabla-\nabla \boldsymbol{U}_b\cdot+Re^{-1}\nabla^2 & -\nabla \\
\nabla\cdot & \mathbf{0} \\
\end{pmatrix}.
\end{equation}

This leads to the eigenvalue problem of matrix $\mathbf{A}$, where the stability of the flow is determined by the sign of the real part of the eigenvalues $\sigma$. The non-trivial solutions of $(\hat{\boldsymbol{u}},\hat{q})^T$ constitute global linear modes of the problem. 

For complex flow configurations (e.g., cylinder flow), it is rarely practical to assemble and store the full matrix representation \(\mathbf{A}\) explicitly. This global eigenvalue problem can be solved by an iterative approach, and the most popular one is the Arnoldi algorithm \citep{arnoldi1951principle,saad1980variations}.  It is a time-stepping–based Jacobian-free method and has been widely used for linear stability analysis in the fluid mechanics community.

\subsection{Orr-Sommerfeld/Squire equations}

For parallel flows, one may simplify the linearized Navier-Stokes equations (\eqref{equ:lin_ns1}) and solve smaller eigenproblems analytically. For instance, as the base flow of the channel flow case is known as $U(y) = 1-y^2$, we can plug it into \eqref{equ:lin_ns1} and obtain its component form:

\begin{align}
\frac{\partial u'}{\partial t} + U\frac{\partial u'}{\partial x} + v'\frac{dU}{dy} &= -\frac{\partial p'}{\partial x} + Re^{-1}\nabla^2 u' \label{equ:lin_x_3d} \\
\frac{\partial v'}{\partial t} + U\frac{\partial v'}{\partial x} &= -\frac{\partial p'}{\partial y} + Re^{-1}\nabla^2 v' \label{equ:lin_y_3d} \\
\frac{\partial w'}{\partial t} + U\frac{\partial w'}{\partial x} &= -\frac{\partial p'}{\partial z} + Re^{-1}\nabla^2 w' \label{equ:lin_z_3d} \\
\frac{\partial u'}{\partial x} + \frac{\partial v'}{\partial y} + \frac{\partial w'}{\partial z} &= 0 \label{equ:lin_cont_3d}
\end{align}

We introduce the wall-normal velocity $v'(x,y,z,t)$ and the wall-normal vorticity $\eta'(x,y,z,t) = \frac{\partial u'}{\partial z} - \frac{\partial w'}{\partial x}$.
By combining \eqref{equ:lin_x_3d}-\eqref{equ:lin_cont_3d} to eliminate the pressure and tangential velocity components, one obtains a coupled system for $v'$ and $\eta'$. The equation governing $v'$ is the three-dimensional Orr-Sommerfeld equation,
\begin{equation}\label{equ:os_3d_physical}
\frac{\partial}{\partial t}(\nabla^2 v') + U \frac{\partial}{\partial x}(\nabla^2 v') - \frac{d^2U}{dy^2}\frac{\partial v'}{\partial x}
= Re^{-1} \nabla^4 v',
\end{equation}
while the wall-normal vorticity satisfies the Squire equation,
\begin{equation}\label{equ:squire_3d_physical}
\frac{\partial \eta'}{\partial t} + U\frac{\partial \eta'}{\partial x} + \frac{dU}{dy}\frac{\partial v'}{\partial z}
= Re^{-1}\nabla^2 \eta'.
\end{equation}
Here, $\nabla^2 = \partial_{xx} + \partial_{yy} + \partial_{zz}$ and $\nabla^4 = \nabla^2\nabla^2$.

We can further assume perturbations in normal modes exhibit both streamwise and spanwise periodic dependence, i.e.
\begin{equation}
\{v',\eta'\}(x,y,z,t)=\{\hat{v}(y),\hat{\eta}(y)\}e^{i(k_x x+k_z z)}e^{\sigma t}
=\{\hat{v}(y),\hat{\eta}(y)\}e^{ik_x(x-ct)}e^{ik_z z},
\end{equation}
where $\hat{v}(y)$ and $\hat{\eta}(y)$ are the wall-normal amplitude functions, $k_x$ and $k_z$ are the real streamwise and spanwise wavenumbers, and $c=c_r+ic_i$ is the complex wave speed. In other words, the perturbation is considered in wavenumber space to facilitate simplification. Substituting this normal mode form into \eqref{equ:os_3d_physical} and \eqref{equ:squire_3d_physical} transforms the partial derivatives as follows:
\[
\frac{\partial}{\partial t}\to -ik_x c,\qquad
\frac{\partial}{\partial x}\to ik_x,\qquad
\frac{\partial}{\partial z}\to ik_z,\qquad
\nabla^2 \to D^2-k^2,
\]
where $D=\frac{d}{dy}$ and $k^2=k_x^2+k_z^2$. After simplification, we arrive at the Orr-Sommerfeld/Squire system:
\begin{align}
(U-c)(D^2-k^2)\hat{v} - \frac{d^2U}{dy^2}\hat{v}
&= \frac{1}{ik_x Re}(D^2-k^2)^2\hat{v}, \label{equ:os_3d_final} \\
(U-c)\hat{\eta} + \frac{k_z}{k_x}\frac{dU}{dy}\hat{v}
&= \frac{1}{ik_x Re}(D^2-k^2)\hat{\eta}. \label{equ:squire_3d_final}
\end{align}

Since the no-slip boundary conditions at the walls ($y=\pm 1$) require $u'=v'=w'=0$, the corresponding boundary conditions become

\[
\hat{v}(\pm 1)=0,\qquad D\hat{v}(\pm 1)=0,\qquad \hat{\eta}(\pm 1)=0.
\]

The Orr-Sommerfeld/Squire equations together with these homogeneous boundary conditions, again form an eigenvalue problem. It is worth noting that the two-dimensional counterpart is recovered by setting $k_z=0$, in which case the Squire equation decouples and the above system reduces to the classical Orr-Sommerfeld eigenvalue problem.

\section{Cyclostationary climate system}

\begin{figure}
\centering
\includegraphics[width=\linewidth]{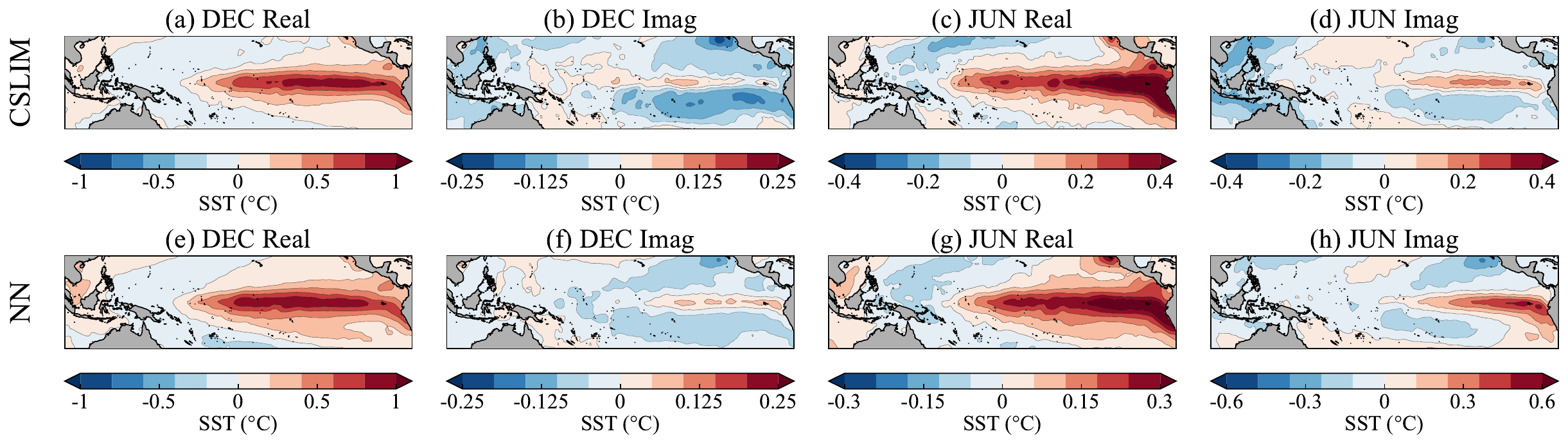}
\caption{Spatial structure of the ENSO SST mode for both December and June predicted by the CSLIM and NN emulator.}\label{fig:sstmode}
\end{figure}

\begin{figure}
\centering
\includegraphics[width=\linewidth]{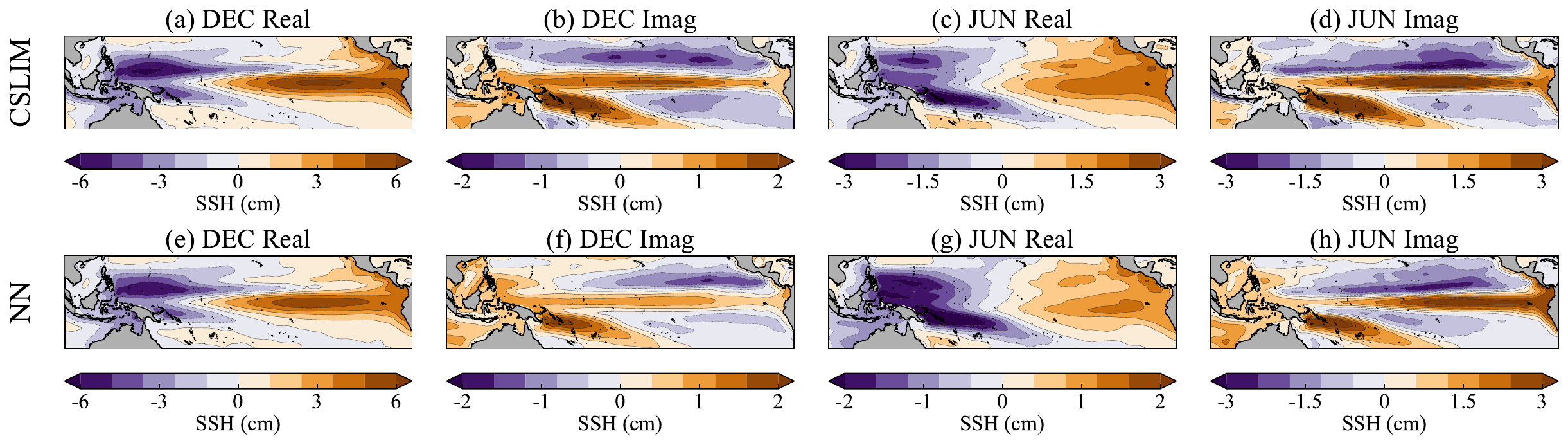}
\caption{Spatial structure of the ENSO SSH mode for both December and June predicted by the CSLIM and NN emulator.}\label{fig:sshmode}
\end{figure}

In reference to Vimont \textit{et al.}~\citep{vimont2022role}, we also illustrate the spatial structures of the ENSO SST/SSH mode for both December and June in Fig.~\ref{fig:sstmode} and Fig.~\ref{fig:sshmode}, including the real and imaginary components. The real component usually depicts the mature phase of ENSO events, and the spatial similarity for the real part between the NN emulator and the CSLIM is exceptionally high. Both are dominated by significant SST warming in the central-eastern equatorial Pacific and corresponding SSH anomalies. The imaginary component is generally more challenging to extract, as it physically represents the phase ahead of the real component by one-quarter of the oscillation time scale. Although the spatial similarity for the imaginary part is slightly lower for both months, it still captures the core features: positive SST and deep thermocline anomalies along the equator, bordered by cool SST and shallow thermocline anomalies on either side of the equator. These spatial structures from both methods are consistent with the theoretical delayed/recharge oscillator paradigm \citep{battisti1988dynamics,jin2008current}.

Note that the amplitude of the NN emulator's ENSO mode decays by nearly 50\% from December to June, whereas the CSLIM mode decays by only about 16\%. This pronounced seasonal decay is internally consistent with the different damping time scales predicted by the two methods (Fig.~9(a) of the main text). The damping time of the ENSO mode in the CSLIM is approximately 19 months, whereas the NN estimate is about 15 months. The faster damping rate exhibited by the NN implies stronger seasonal modulation, leading to a more pronounced and realistic winter-peaked characteristic \citep{timmermann2018nino}. 

In \citep{vimont2022role}, CSLIM employs the 3-month lag covariance matrix to ensure the robustness of the parameter estimation in fitting the monthly propagators. This stationary linear approximation inevitably introduces a time-averaging effect on the seasonal evolution of the mode, thereby smoothing out its intrinsically sharp winter-peaked characteristics. In contrast, the NN emulator directly learns the state-dependent nonlinear dynamics. However, a drawback is the large error bars observed in the instantaneous growth rate curves (Fig.~9(b) of the main text), indicating that neural networks trained with different random seeds are sensitive to small fluctuations in the samples when estimating local instantaneous gradients.

Overall, although the NN emulator suffers from statistical uncertainties in extracting the system matrix gradients, the stronger winter phase-locking and decay characteristics it reveals in the ENSO mode provide an alternative physical perspective on the seasonal evolution, which might otherwise be obscured by the smoothing assumptions inherent in traditional statistical linear models. The CSLIM and the NN are essentially two effective approximations of the same cyclostationary ENSO dynamics from different methodological standpoints. Nevertheless, which of these two portrayals more closely resembles the real climate system remains an open question that warrants further investigation through controlled experiments.

\section{Additional examples}

\subsection{Lorenz system}

To show the capacity of our method on more chaotic systems, we test on the Lorenz system. NNs have been proven to be capable of characterizing its temporal dynamics, including accurately advancing the solution in time for unseen initial conditions \citep{brunton2022data}. The state of the Lorenz system given by $\boldsymbol{q} = [x, y, z]^T$ is governed by

\begin{equation}
\begin{aligned}
      \frac{\mathrm{d}x}{\mathrm{d}t} &= \sigma\left(y-x\right), \\
      \frac{\mathrm{d}y}{\mathrm{d}t} &= x\left(\rho-z\right)-y, \\
      \frac{\mathrm{d}z}{\mathrm{d}t} &= xy-\beta z.
\end{aligned}
\end{equation}

\begin{figure}
  \centerline{\includegraphics[width=0.9\linewidth]{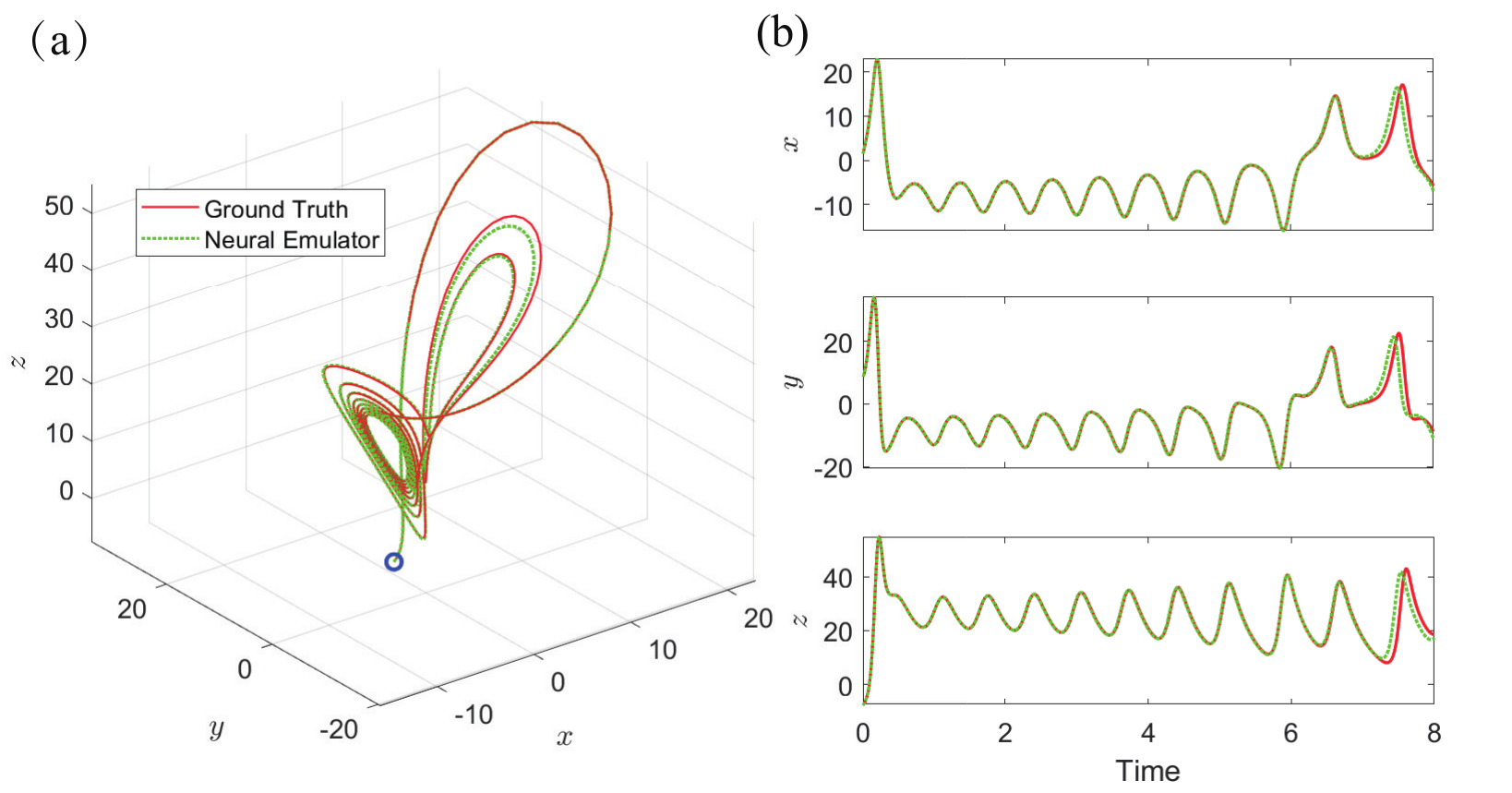}}
  \caption{Time evolution of the true Lorenz system (red solid) compared with the NN emulator prediction (green dotted) from an unseen initial point (blue circle). (a) Trajectories in phase space; (b) Component-wise time evolution series.}
  \label{fig:lorenz}
\end{figure}

\begin{sidewaystable}
  \centering
  \begin{tabular}{cccccc} \toprule
                      \multicolumn{2}{c}{\multirow{2}{*}{Base state $\boldsymbol{q}_b$}}  & \multicolumn{2}{c}{NN-based Results} & \multicolumn{2}{c}{Analytical Results} \\ \cline{3-6} \addlinespace[2pt]
                      \multicolumn{2}{c}{}  & Jacobian $\mathbf{N}$         & eigenvalues         & Jacobian $\exp(\mathbf{A} \Delta t)$         & eigenvalues         \\ \midrule \addlinespace[5pt]
                      & $\begin{bmatrix} 0 \\ 0 \\ 0 \end{bmatrix}$ & $\begin{bmatrix} 0.9185 & 0.0947 & -0.0001 \\ 0.2657 & 1.0036 &  0.0005 \\ 0.0002 & -0.0001 &  0.9735 \end{bmatrix}$    & $\begin{bmatrix} 1.1253 \\ 0.9735 \\ 0.7968 \end{bmatrix}$ & $\begin{bmatrix} 0.9179 & 0.0951 & 0 \\ 0.2663 & 1.0035 & 0 \\ 0 & 0 & 0.9737 \end{bmatrix}$                 & $\begin{bmatrix} 1.1256 \\ 0.9737 \\ 0.7959 \end{bmatrix}$                  \\ \addlinespace[5pt]
equilibrium   & $\begin{bmatrix} 6\sqrt{2} \\ 6\sqrt{2} \\ 27 \end{bmatrix}$  & $\begin{bmatrix} 0.9054 & 0.0945 & -0.0041 \\ 0.0074 & 0.9861 & -0.0836 \\ 0.0789 & 0.0880 & 0.9705 \end{bmatrix}$                & $\begin{bmatrix} 0.8706 \\ 0.9957\pm0.1018i \end{bmatrix}$ & $\begin{bmatrix} 0.9052 & 0.0946 & -0.0041 \\ 0.0060 & 0.9869 & -0.0832 \\ 0.0800 & 0.0873 & 0.9700 \end{bmatrix}$                 & $\begin{bmatrix} 0.8706 \\ 0.9957\pm 0.1019i \end{bmatrix}$                  \\ \addlinespace[5pt]
                      & $\begin{bmatrix} -6\sqrt{2} \\ -6\sqrt{2} \\ 27 \end{bmatrix}$ & $\begin{bmatrix} 0.9050 & 0.0948 & 0.0039 \\ 0.0071 & 0.9865 & 0.0835 \\ -0.0812 & -0.0869 & 0.9697 \end{bmatrix}$                & $\begin{bmatrix} 0.8694 \\ 0.9959\pm0.1019i \end{bmatrix}$ & $\begin{bmatrix} 0.9052 & 0.0946 & 0.0041 \\ 0.0060 & 0.9869 & 0.0832 \\ -0.0800 & -0.0873 & 0.9700 \end{bmatrix}$ & $\begin{bmatrix} 0.8706 \\ 0.9957\pm 0.1019i \end{bmatrix}$                \\ \addlinespace[5pt] \midrule \addlinespace[5pt]
non-equilibrium & $\begin{bmatrix} 1.4689 \\ 8.7045 \\ -7.6905 \end{bmatrix}$ & $\begin{bmatrix} 0.9221 & 0.0941 & 0.0001 \\ 0.3388 & 1.0075 & -0.0180 \\ 0.0855 & 0.0238 & 0.9734 \end{bmatrix}$ & $\begin{bmatrix} 1.1446 \\ 0.9784 \\ 0.7800 \end{bmatrix}$ & $\begin{bmatrix} 0.9215 & 0.0952 & -0.0007 \\ 0.3393 & 1.0071 & -0.0145 \\ 0.0847 & 0.0187 & 0.9736 \end{bmatrix}$ & $\begin{bmatrix} 1.1461 \\ 0.9775 \\ 0.7786 \end{bmatrix}$ \\ \addlinespace[5pt] \bottomrule
\end{tabular}
  \caption{Comparison of the NN-based Jacobian matrix $\mathbf{N}$ and the operator-based Jacobian matrix $\mathbf{A}$ around three equilibrium points and one non-equilibrium point (the initial point in Fig.~\ref{fig:lorenz}). }
  \label{tab:lorenz}
\end{sidewaystable}

This system has three equilibrium points, one at the origin ($[0, 0, 0]^T$) and two symmetric off-origin points for $\rho > 1$ ($[\pm\sqrt{\beta(\rho-1)}, \pm\sqrt{\beta(\rho-1)}, \rho-1]$). By applying a local linearization to the Lorenz system around a base state $\boldsymbol{q}_b=[x_b,y_b,z_b]^T$, the analytical local Jacobian is obtained:

\begin{equation}\label{equ:lorenz}
\mathbf{A} = \frac{\partial \mathcal{N}}{\partial \boldsymbol{q}}\Big|_{\boldsymbol{q}_b} =
\begin{bmatrix}
-\sigma & \sigma & 0 \\
\rho-z_b & -1 & -x_b \\
y_b & x_b & -\beta
\end{bmatrix}.
\end{equation}

Here, the standard parameter values of $\sigma = 10$, $\rho = 28$, and $\beta = 8/3$ are chosen, and thus the system exhibits chaotic motion confined to a strange attractor. The training data is constructed from high-fidelity simulations of the Lorenz system via the fourth-order Runge–Kutta method. $100$ different initial conditions are advanced in time for a total of $m=800$ snapshots with a fixed $0.01$ time units. The NN architecture is a three-layer network with 10 neurons per hidden layer adopted from \citep{brunton2022data}.
And a one-step rollout strategy is applied.

Similar to the mean-field model in the main text, Fig.~\ref{fig:lorenz}(a) presents the trajectory comparison between the NN emulator's prediction and ground truth, starting from a randomly selected initial condition over an interval of 8 time units. Fig.~\ref{fig:lorenz}(b) offers a clearer view of the approximation's performance by showing the time evolution of three components separately. Due to the intrinsic sensitivity of the Lorenz system to initial perturbations, the two simulations show close agreement in the early stage, while a noticeable deviation is observed for the NN emulator at the end of the simulation. Even so, the performance of the NN emulator remains commendable given the lack of rollout steps at training time. In Tab.~\ref{tab:lorenz}, we compare the NN-based Jacobians around the three equilibrium points and one non-equilibrium point with the analytical Jacobians derived from Eq.~\ref{equ:lorenz}. The NN‑based Jacobians closely match their analytical counterparts, with nearly identical eigenvalues, as anticipated.

\subsection{Reduced complex Ginzburg-Landau system}

To further verify the spectral consistency between the POD-reduced system and the original physical-space system, we here consider a POD-reduced experiment for the complex Ginzburg-Landau system in the main text. The sample trajectories and all parameter settings are kept identical to those of the full-state case, except that here we consider only the nonlinear dataset. Since the original spatial discretization yields a state vector of dimension $220$, we retain only the first $20$ POD modes in the reduced representation. A four-layer complex-valued perceptron with 256 neurons in each hidden layer is employed, and only a ten-step rollout is used during training.

\begin{figure}
\centering
\includegraphics[width=\linewidth]{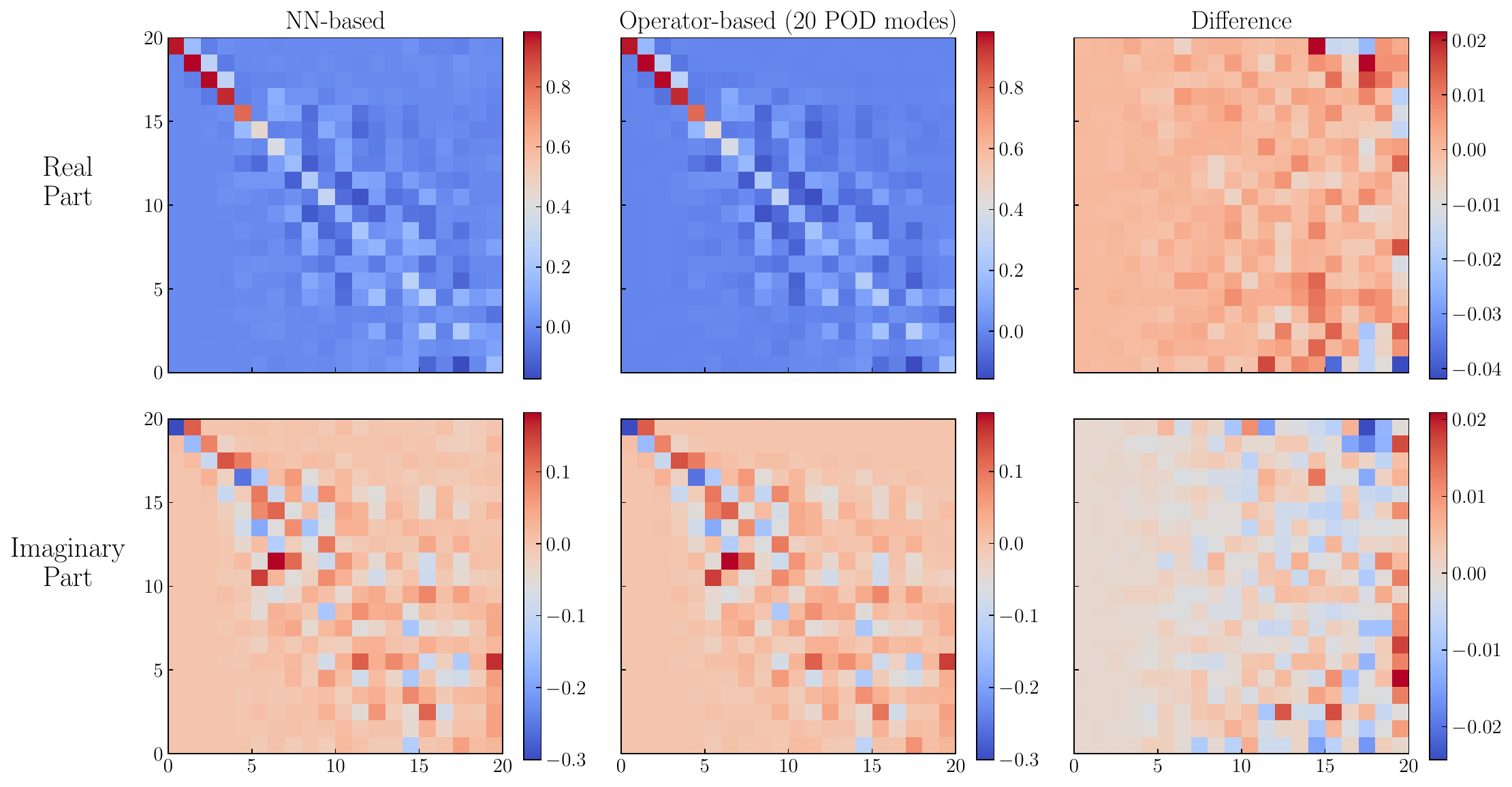}
\caption{Comparison of the real and imaginary parts of the reduced Jacobians between the NN-based operator and the operator-based ground truth.}\label{fig:reduced_jacobian}
\end{figure}

As shown in Fig.~\ref{fig:reduced_jacobian}, we compare the heatmaps of the NN-based Jacobian learned from the reduced data with the analytical counterpart $\mathrm{N}_{\mathrm{POD}}$ obtained by projecting the full analytical Jacobian onto the POD subspace. The improvement is particularly evident from the prediction difference in Fig.~\ref{fig:reduced_jacobian}, where the error magnitude is significantly smaller than that of the corresponding full-order case shown in the main text. This improvement is expected since POD substantially reduces the input and output dimensions of the neural emulator, thereby simplifying the learning problem. 

\begin{figure}
\centering
\includegraphics[width=0.9\linewidth]{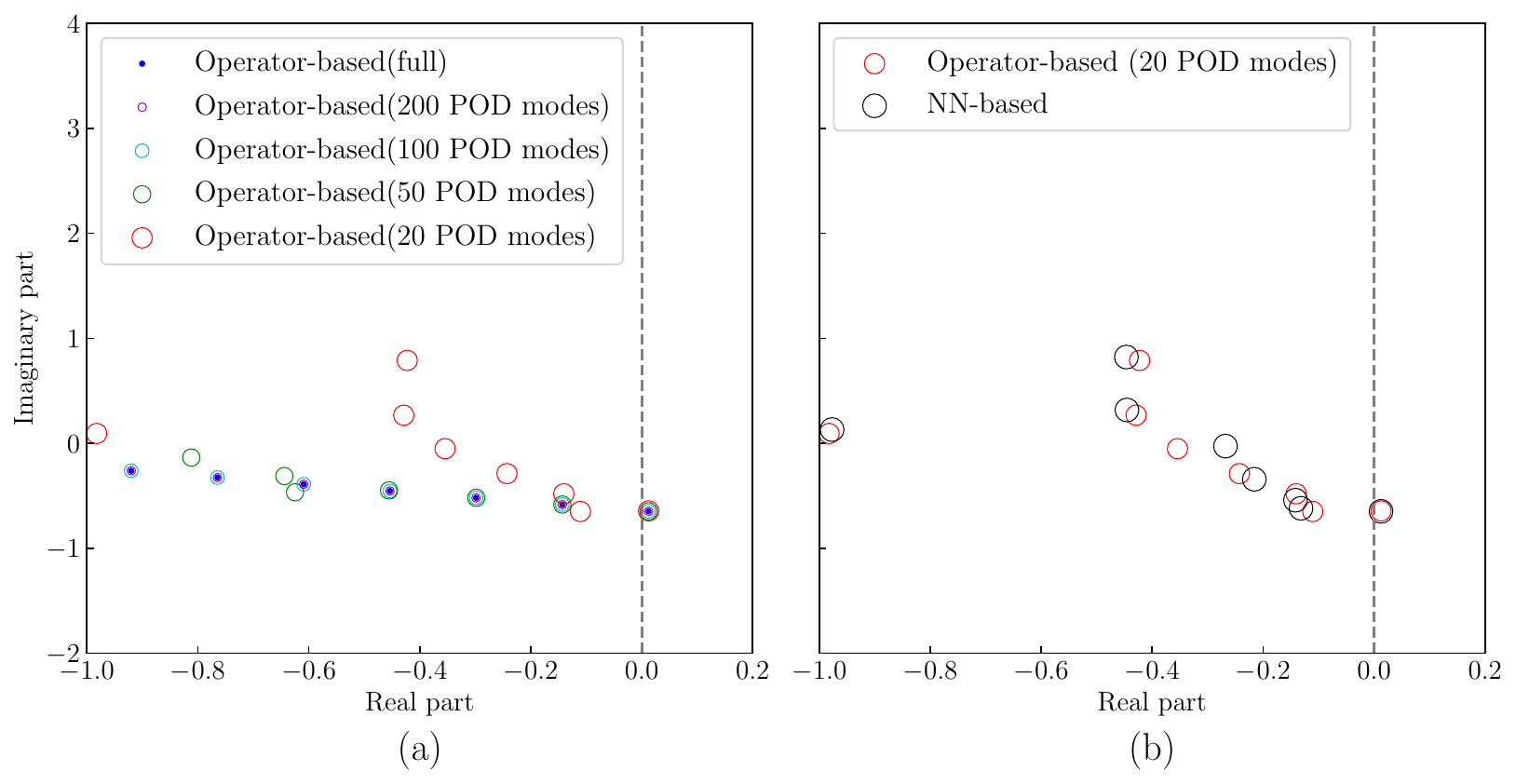}
\caption{Linear stability analysis results of the POD-reduced complex Ginzburg-Landau systems. (a) Operator-based eigenvalues obtained by projecting the full Jacobian onto POD subspaces with different truncation levels, compared with the full-order spectrum. (b) Comparison between the operator-based and NN-based eigenspectra in the 20-mode POD subspace.}
\label{fig:reduced_eigval}
\end{figure}

As we mentioned in the main text, if one is only interested in the leading eigenvalues, it is not necessary to retain all POD modes to construct this subspace. In Fig.~\ref{fig:reduced_eigval}(a), we show the consistency between the eigenvalues of the full Jacobian $\mathbf{N}$ and those of $\mathbf{N}_{\mathrm{POD}}$ when different numbers of POD modes are retained. It can be seen that 100 leading POD modes are already sufficient to recover the first seven eigenvalues with high accuracy. Even with only 20 POD modes, the reduced system still captures the unique leading unstable mode of the system.

Figure~\ref{fig:reduced_eigval}(b) further shows the eigenvalues predicted by the neural emulator trained on the reduced POD coefficients. In addition to the leading unstable mode, many higher-order modes are also found to approach the analytical spectrum. This observation is consistent with the very small Jacobian error seen in Fig.~\ref{fig:reduced_jacobian}. By contrast, the NN-based eigenvalues obtained from training on the full-order data show much poorer accuracy for the higher-order modes (Fig.~4(b) in the main text).

This behavior is physically reasonable for the complex Ginzburg-Landau system. For most snapshots, the state amplitude is concentrated only in a localized region near the origin, while the far-field state remains close to zero (Fig.~\ref{fig:cgle}). Consequently, when training directly on the full-order data, a large fraction of the input-output mapping learned by the neural network is spent on dynamically uninformative regions of the domain. By projecting the state onto a compact subspace spanned by the dominant POD modes, the neural emulator is relieved from learning spatially redundant information and can focus on the temporal‑coefficient dynamics associated with structures at different energy levels. As a consequence, the resulting reduced-order model yields a more accurate and significantly improved prediction of the Jacobian compared with direct full-state learning.

\subsection{2D transitional channel flow}

We then consider the two-dimensional version of the channel flow case with $Re=2000$, but now in the original physical state rather than the reduced space used for the 3D case in the main text. Owing to the smaller spatial grid, the Jacobian obtained via automatic differentiation is of modest size, allowing us to directly demonstrate the NN‑based modal analysis in physical space. This lightweight setting also enables us to further probe how the strength of nonlinearity in the dataset affects the results.

\begin{figure}
\centering
\includegraphics[width=0.8\linewidth]{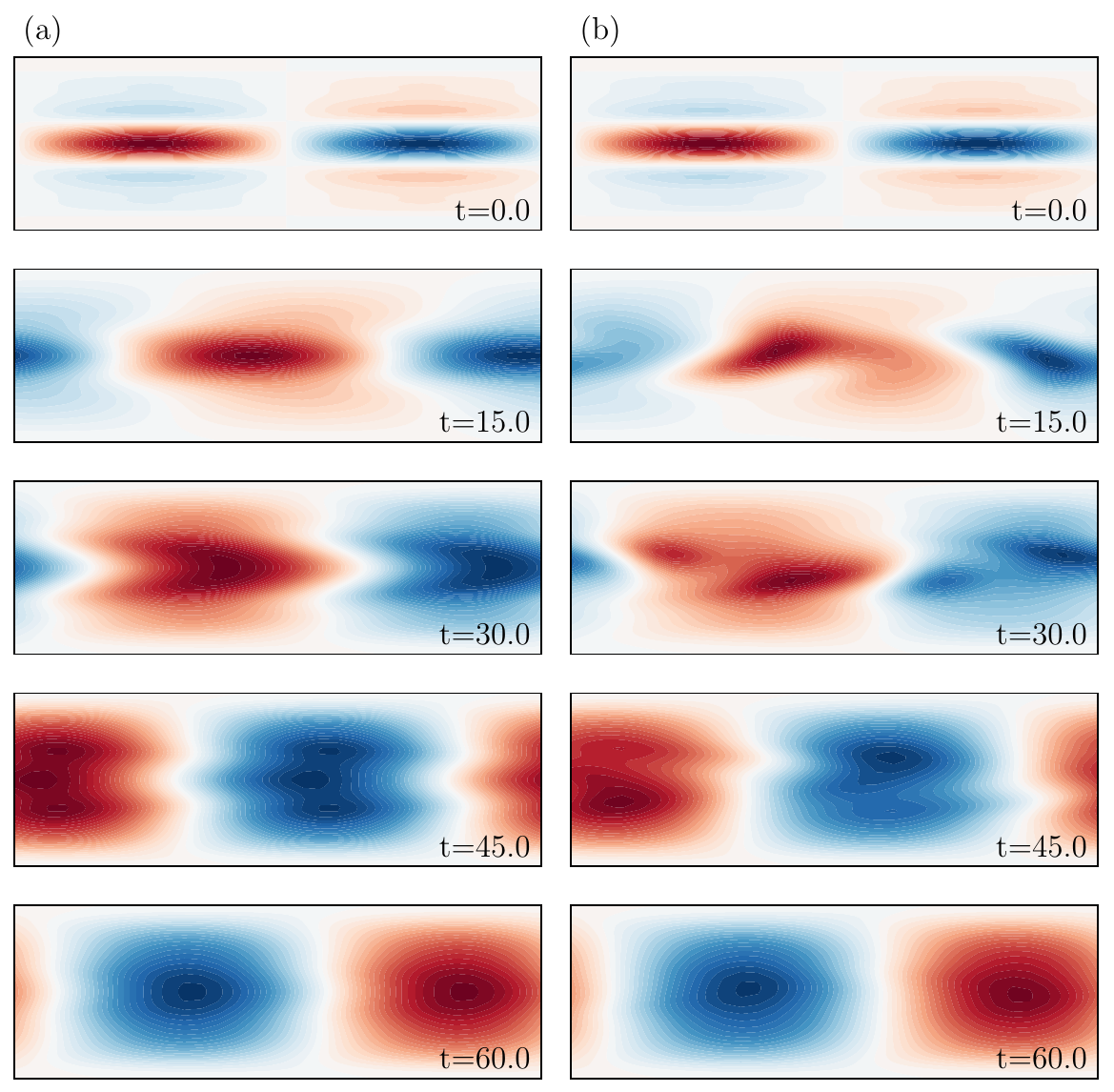}
\caption{Snapshots of the wall-normal velocity perturbation field during the transient evolution stage of two-dimensional channel flow. The initial perturbation profiles are the same but with different magnitude scaling factors $\epsilon$: (a) weakly nonlinear scenario with $\epsilon = 10^{-6}$; (b) nonlinear scenario with $\epsilon = 10^{-1}$. The decaying perturbations in (a) remain symmetric structures when evolving, while those in (b) encounter symmetry-breaking phenomena.}
\label{fig:chan_snap}
\end{figure}

The overall setup mirrors the three-dimensional case in the main text. The main difference is that the spanwise direction is removed, and the perturbations are advanced at a single streamwise wavenumber $k_x = 1$ with a time step of $0.002$. Snapshots are collected during the transient growth stage, with $60$ trajectories and $m = 400$ snapshots saved every $0.5$ time units. Accordingly, the initial velocity perturbation fields simplify to

\begin{equation}
\begin{aligned}
u(x,y) &= \epsilon \cdot \Re\left[\phi'(y)\, e^{\mathrm{i}k_x x}\right], \\
v(x,y) &= -\epsilon \cdot \Re\left[\mathrm{i}\,\phi(y)\, e^{\mathrm{i}k_x x}\right],
\end{aligned}
\end{equation}

\noindent where $\phi(y)$ is constructed as in the main text and $\epsilon$ scales the initial perturbation magnitude, thereby setting the strength of the nonlinear interactions. Its pronounced influence on the transient flow behavior is illustrated in Fig.~\ref{fig:chan_snap}. In contrast to the single strongly nonlinear dataset used for the 3D case, here we generate both a weakly nonlinear dataset with $\epsilon = 10^{-6}$ and a strongly nonlinear one with $\epsilon = 10^{-1}$, which lets us isolate the effect of nonlinearity on the learned operator.

Because the state is retained in physical space, we learn the flow map with a classic U-Net architecture \citep{ronneberger2015u} with four encoding and decoding layers and $32$ channels in the top layer, rather than the multilayer perceptron acting on POD coefficients employed for the 3D case. For the resolvent analysis, a truncation level of $r = 30$ is used for the eigenbasis of the projection subspace, and ten steps of temporal unrolling are used during training.

\begin{figure}
  \centerline{\includegraphics[width=\linewidth]{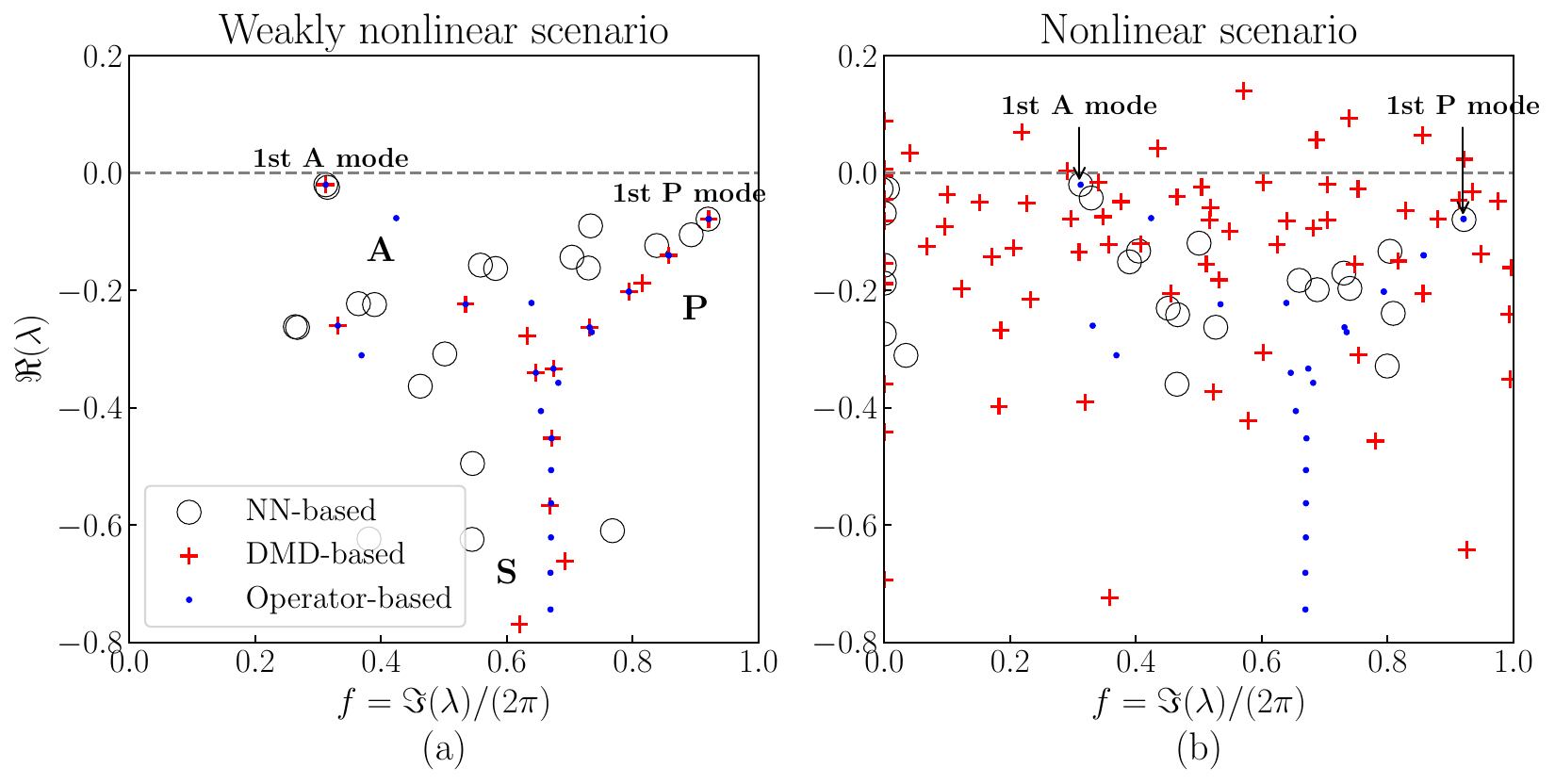}}
  \caption{Comparison of the eigenspectra obtained from the NN-based Jacobian ($\circ$), DMD-based Jacobian ({\color{red}$+$}) and operator-based ground truth ($\color{blue}\bullet$) for the 2D channel flow system. The dashed line denotes the stability boundary in the complex plane. (a) Weakly nonlinear dataset. (b) Nonlinear dataset.}
  \label{fig:chan_eigenspectrum}
\end{figure}

\begin{figure}
  \centerline{\includegraphics[width=\linewidth]{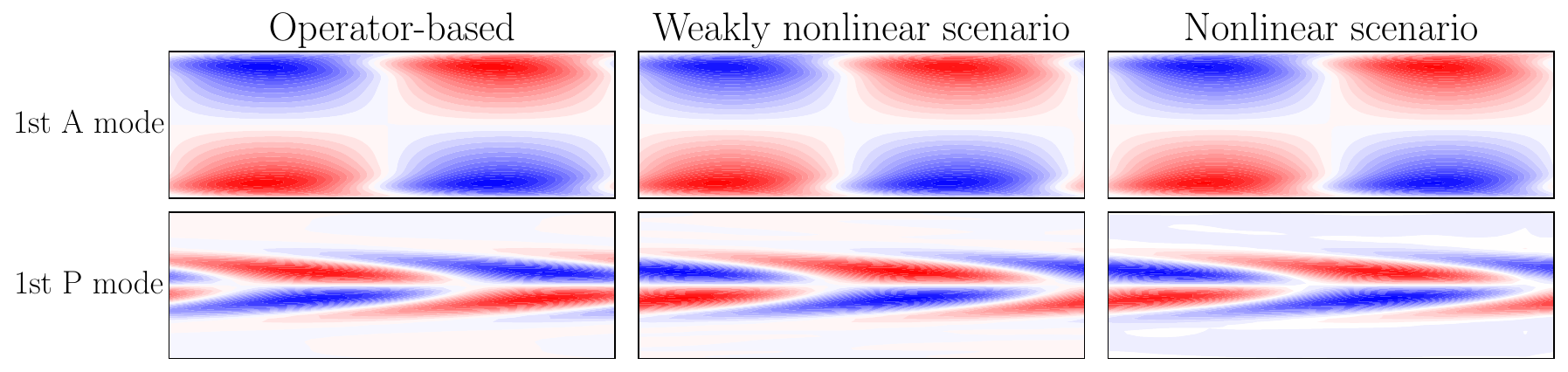}}
  \caption{Comparison of the first-order A and P eigenmodes of the streamwise velocity field for the 2D channel flow system. 
  The first column is the reference results obtained from the OS operator. The second and third columns are the NN-based Jacobian obtained from weakly nonlinear and nonlinear datasets.}
  \label{fig:chan_eigenvector}
\end{figure}

Fig.~\ref{fig:chan_eigenspectrum}(a) and (b) compare the NN-based, DMD-based, and operator-based eigenspectra for the weakly and strongly nonlinear datasets, respectively, with the corresponding leading A and P eigenmodes shown in Fig.~\ref{fig:chan_eigenvector}. All eigenvalues again lie in the stable half-plane and organize into the Y-shaped A-, P-, and S-branches described in the main text. 

For our approach, the leading eigenvalues of the A- and P-branches are recovered accurately for both datasets, confirming that the NN emulator identifies the two most dominant near-wall and center structures even under strong nonlinearity. Consistent with the two spatial dimensions, these modes appear as simple wave packets propagating along the streamwise direction (Fig.~\ref{fig:chan_eigenvector}). Beyond the leading modes, however, training on the full physical state does not reproduce the branch structure. The higher-order A and P eigenvalues in Fig.~\ref{fig:chan_eigenspectrum}(b) are largely spurious. This is precisely the behavior contrasted in the main text, where learning the dynamics within the POD subspace instead allows several secondary A and P modes to fall along the analytical branches. Nevertheless, our approach successfully determines the correct stability characteristics of the system, while the randomly dispersed DMD eigenvalues in Fig.~\ref{fig:chan_eigenspectrum}(b) further highlight the failure of a purely linear model when applied to strongly nonlinear data.

The two datasets also make the preprocessing artifact discussed in the main text directly visible. For the strongly nonlinear dataset, additional spurious eigenvalues predicted by the NN emulator appear on the imaginary axis in Fig.~\ref{fig:chan_eigenspectrum}(b), which are absent for the weakly nonlinear dataset in Fig.~\ref{fig:chan_eigenspectrum}(a). The clean weakly nonlinear spectrum confirms that the artifact is a direct consequence of the nonlinearity strength.

\begin{figure}
  \centerline{\includegraphics[width=\linewidth]{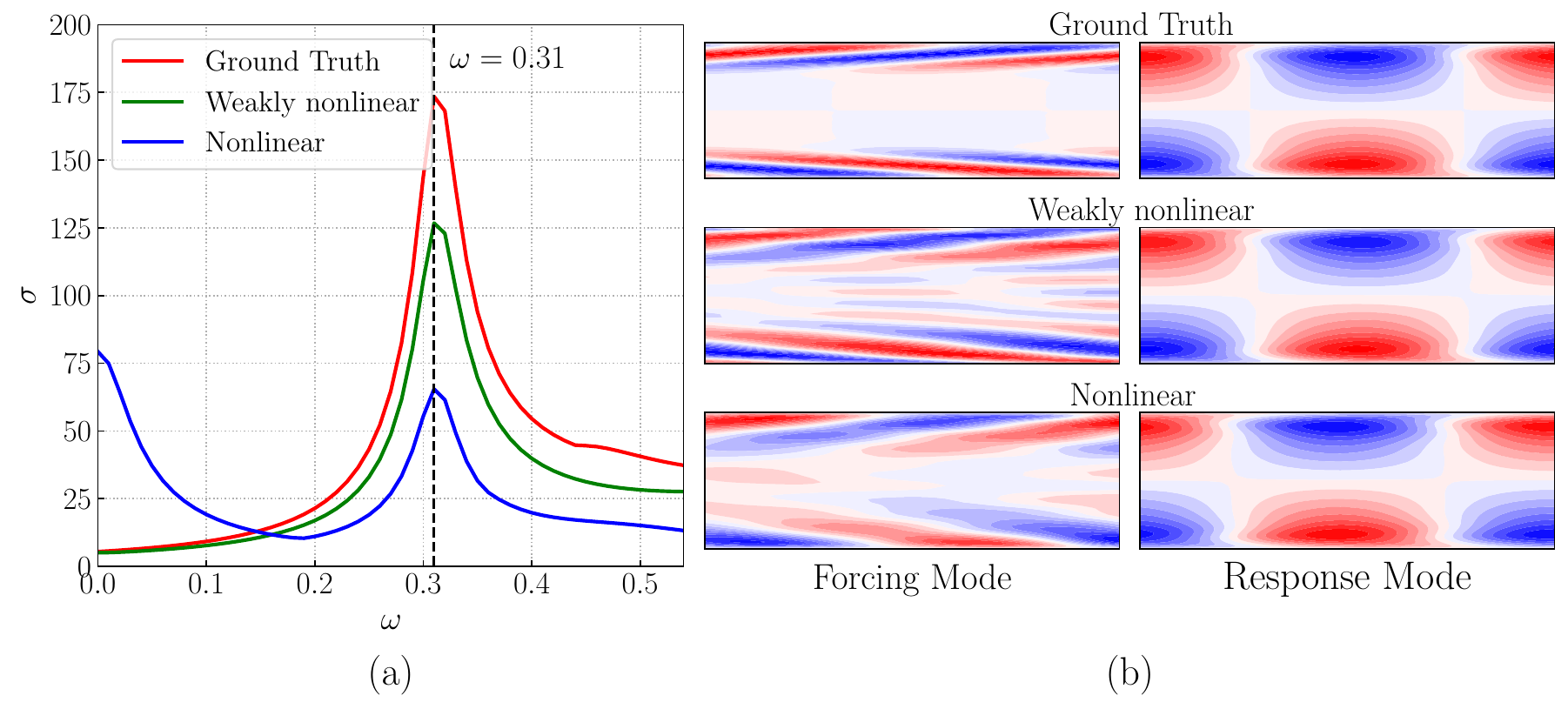}}
  \caption{Resolvent analysis results of the 2D channel flow system. (a) Resolvent gain curves of the analytical operator and the NN-based resolvent operators derived from weakly nonlinear and nonlinear datasets. (b) The first-order forcing and response modes of the streamwise velocity field at the $\omega=0.31$.}
  \label{fig:chan_resolvent}
\end{figure}

For the resolvent analysis, Fig.~\ref{fig:chan_resolvent}(a) shows the gain peak now shifts to $\omega = 0.31$. Mirroring the 3D case, the strongly nonlinear gain curve exhibits an additional peak at zero frequency, which stems from resonance with the non-oscillatory spurious modes. The corresponding forcing and response modes at $\omega = 0.31$ are shown in Fig.~\ref{fig:chan_resolvent}(b). As in the Ginzburg-Landau case, the response mode agrees almost exactly with the reference, whereas the forcing mode is less accurate owing to the non-normality of the operator. This gap widens for the strongly nonlinear dataset, indicating that nonlinearity further complicates training under the present architecture.

\subsection{2D cylinder flow}

To demonstrate a complex flow configuration, we consider a flow with an immersed obstacle: the two-dimensional cylinder flow at Re = 100 (based on the cylinder diameter and inflow velocity). This Reynolds number setting is greater than the critical threshold, and the cylinder wake flow experiences a Hopf bifurcation and evolves to be a time-periodic non-symmetric state. In contrast to channel flow, the linearized operator for such an intricate configuration cannot be derived analytically and must instead be computed numerically. The spectral element method \citep{osti_code-54678,jfm:loiseau:2014} is used for the steady-state solution, numerical linear stability analysis, and fully nonlinear simulations. 

The cylinder with a diameter $D=1$ is placed at $(0, 0)$ in a grid extending from $-15D$ to $35D$ and $-15D$ to $15D$ in $x$ and $y$ directions, respectively. We employ the same boundary conditions as in \citep{barkley2006linear} for linearized and nonlinear NS equations, respectively. The nonlinear simulations are initialized with the superposition of the base flow and different infinitesimal perturbations for faster saturation. The flow snapshots are uniformly resampled within $[-8, 12] \times [-3, 3]$ using an $80 \times 24$ Cartesian grid. A masking approach is applied by assigning zero values to the network inputs and outputs inside the cylinder. The dataset comprises 30 trajectories, each sampled for $m=200$ snapshots every $0.3$ time units. A third-order temporal scheme with a time step of $0.005$ is chosen for nonlinear numerical simulation. The network architecture is similar to that of the 2D channel flow case, except that the input tensor now comprises three features: pressure, streamwise velocity, and transverse velocity. The projection subspace is truncated to $r=30$ modes, and the model is trained using a twenty-step unrolling strategy to ensure coverage of a full period of vortex shedding.

\begin{figure}
  \centerline{\includegraphics[width=\linewidth]{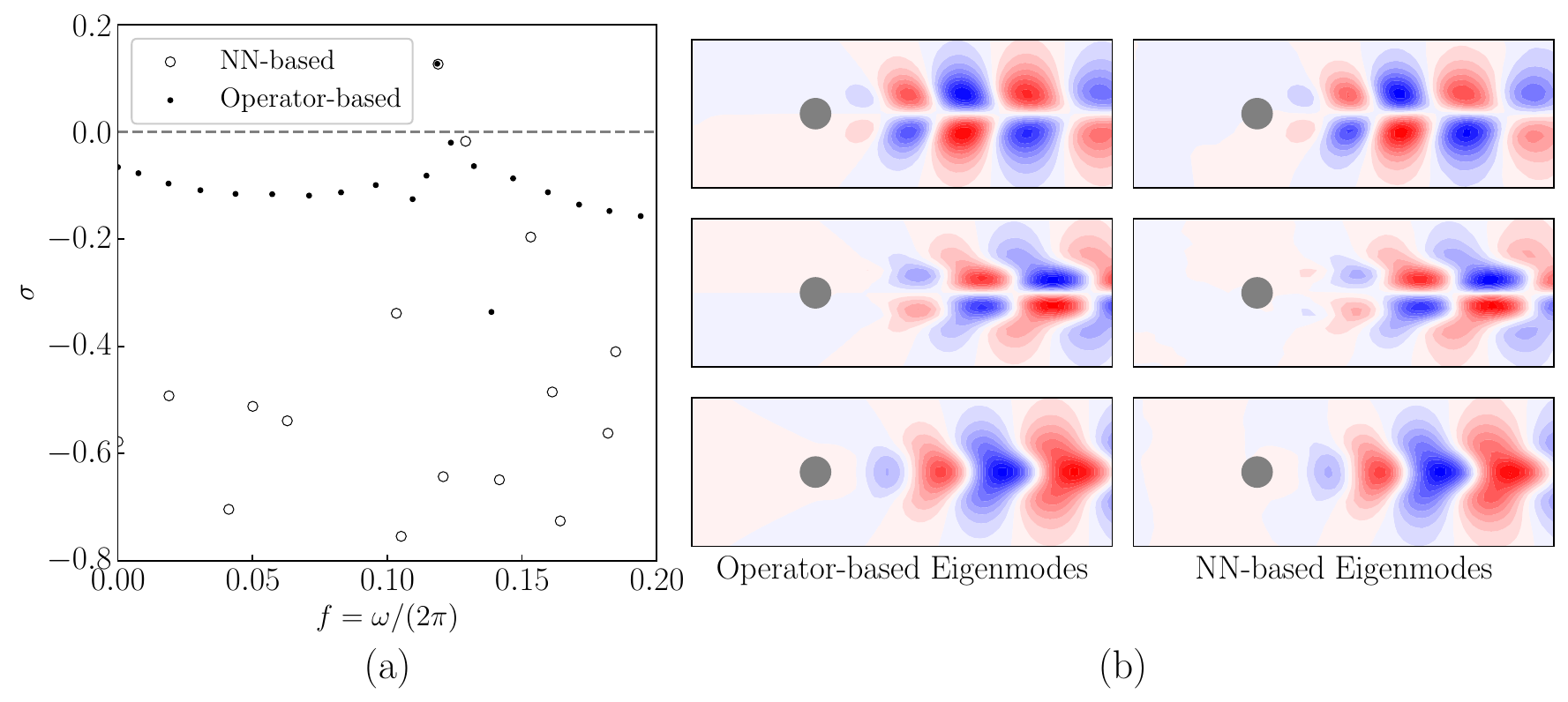}}
  \caption{Linear stability analysis results of the 2D cylinder flow system.  (a) Eigenspectra of the numerical direct stability analysis (operator-based) and NN-based Jacobian for unsteady two-dimensional cylinder flow. (b) Eigenmodes of the unstable eigenvalue from the operator-based and NN-based Jacobians. From top to bottom: pressure, streamwise velocity, and transverse velocity.}
  \label{fig:cyl_eigenspectrum}
\end{figure}

\begin{figure}
  \centerline{\includegraphics[width=0.85\linewidth]{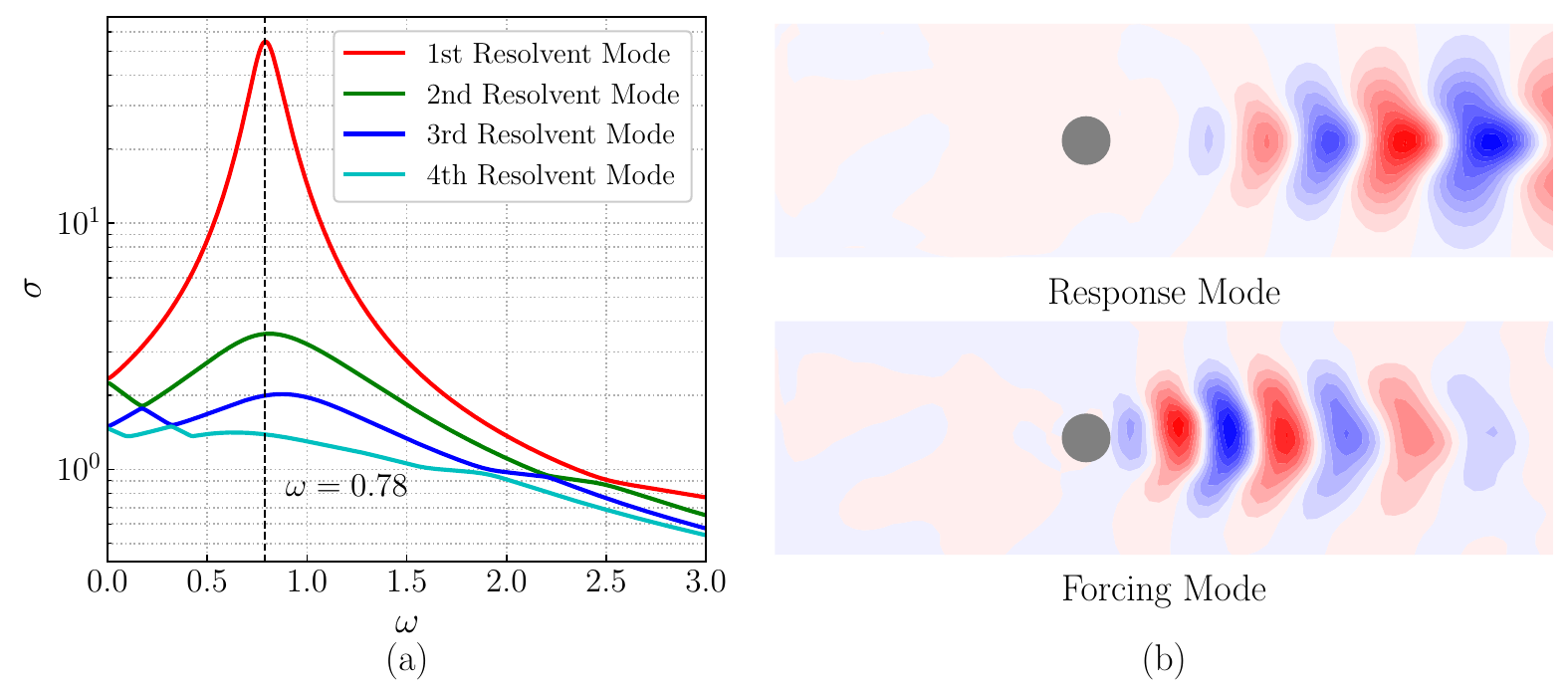}}
  \caption{Resolvent analysis results of the 2D cylinder flow system. (a) Resolvent gain as a function of frequency in the interval [0, 3]. (b) Optimal pressure forcing and response modes evaluated at the frequency corresponding to the peak gain.}
  \label{fig:cyl_resolvent}
\end{figure}

The eigenspectra obtained from numerical linear stability analysis and NN-based Jacobian are presented in Fig.~\ref{fig:cyl_eigenspectrum}(a), along with the leading eigenmodes shown in Fig.~\ref{fig:cyl_eigenspectrum}(b). As expected, the cylinder flow at $Re = 100$ exhibits one unstable leading eigenvalue located in the upper half of the complex plane. The leading eigenvalue predicted by the NN emulator at frequency $f=0.12$ aligns well with the numerical ground truth, and the spatial structures in the downstream area of the cylinder in the corresponding eigenmode also show an excellent agreement. This observation again demonstrates that the NN emulator can not only advance the flow field in an autoregressive fashion, but also capture spatial perturbation structures consistent with those in the numerical modal analysis, thereby reflecting the underlying dynamics of the system.

Note that the frequency of the leading mode differs from the vortex shedding frequency, i.e., Strouhal number ($0.166$) at $Re=100$. Previous literature \citep{barkley2006linear} has reported that only linear stability analysis around the mean flow can predict the vortex shedding frequency with good accuracy, but we reiterate that our current method is not applicable to time-averaged states, as time-averaged states generally do not satisfy the governing equations. Since the steady-state solution of cylinder flow at $Re=100$ represents an artificial pseudo-equilibrium that is not physically realizable, any perturbation introduced around this base state will grow exponentially, ultimately leading to vortex shedding. Consequently, there is no reference for resolvent analysis based on this steady-state solution. In prior resolvent studies of cylinder wakes, the time-averaged flow is typically adopted as the base state, with the unknown nonlinear terms treated as effective forcing to uncover coherent structures.

Nevertheless, we can follow the procedure of previous examples to conduct the resolvent analysis around the steady-state flow based on the obtained NN-based Jacobian. The resulting gain curves and resolvent pressure modes at the optimal forcing frequency are presented in Fig.~\ref{fig:cyl_resolvent}. 
Despite the lack of a ground truth solution, the results offer some clear physical insights: the optimal forcing frequency again coincides with the resonance frequency in Fig.~\ref{fig:cyl_eigenspectrum}, and the corresponding response mode forms symmetric lobe-like patterns downstream of the cylinder, similar to the leading eigenmode. The forcing mode also displays lobe-like structures in the downstream region much closer to the cylinder. This is likely associated with the wavemaker region, the area immediately downstream of the cylinder that is most sensitive to external forcing \citep{giannetti2007structural, marquet2008sensitivity}.

\section{Ablation study}

In this section, we present a series of ablation and robustness studies based on the Lorenz system to further improve our understanding of neural emulators. Before doing so, we first introduce the metrics used to evaluate emulator performance.

To quantify the emulator prediction accuracy, we adopt the metrics introduced in our previous work \citep{koehler2024apebench}, namely the normalized root mean squared error (nRMSE) and its temporally aggregated version. The mean nRMSE over $M$ samples (30 by default) is defined as
\begin{equation}
L_{\text{nRMSE}} = \frac{1}{M} \sum_{i=1}^{M} \sqrt{\frac{ \left\| \bar{\boldsymbol{q}}_i - \boldsymbol{q}_i \right\|^2 }{\left\| \boldsymbol{q}_i \right\|^2}}.
\end{equation}

This metric accounts for scale differences between the predicted state $\bar{\boldsymbol{q}}$ and the reference state $\boldsymbol{q}$, which is especially useful when evaluating rollout errors for dynamical systems whose state amplitudes vary substantially in time, such as decaying channel flows.

Since we are particularly interested in how the error develops over time, we further aggregate the nRMSE over 100 rollout steps using the geometric mean (Agg. nRMSE)
\begin{equation}
L_{\text{Agg. nRMSE}} = \exp \left( \frac{1}{100} \sum_{t=1}^{100} \log(L_{\text{nRMSE}}^{[t]}) \right),
\end{equation}
where the superscript $[t]$ denotes the nRMSE after $t$ time steps. For the Lorenz system, whose Lyapunov time is of order one time unit, this 100-step aggregation provides a meaningful measure of temporal generalization. The geometric mean removes the requirement to hand-tune upper limits for temporal aggregation in cases where error metrics surpass 1.

The relative Frobenius-norm error is used to quantify the prediction accuracy of element-wise Jacobian evaluated at state $\boldsymbol{q}$ as follows:

\begin{equation}
    \varepsilon(\boldsymbol{q})=
\frac{\left\| \mathbf{J}_{\mathrm{NN}}(\boldsymbol{q})-\mathbf{J}_{\mathrm{ana}}(\boldsymbol{q}) \right\|_{F}}
{\left\| \mathbf{J}_{\mathrm{ana}}(\boldsymbol{q}) \right\|_{F}}.
\end{equation}

\subsection{Dataset range}
\begin{figure}
  \centerline{\includegraphics[width=\linewidth]{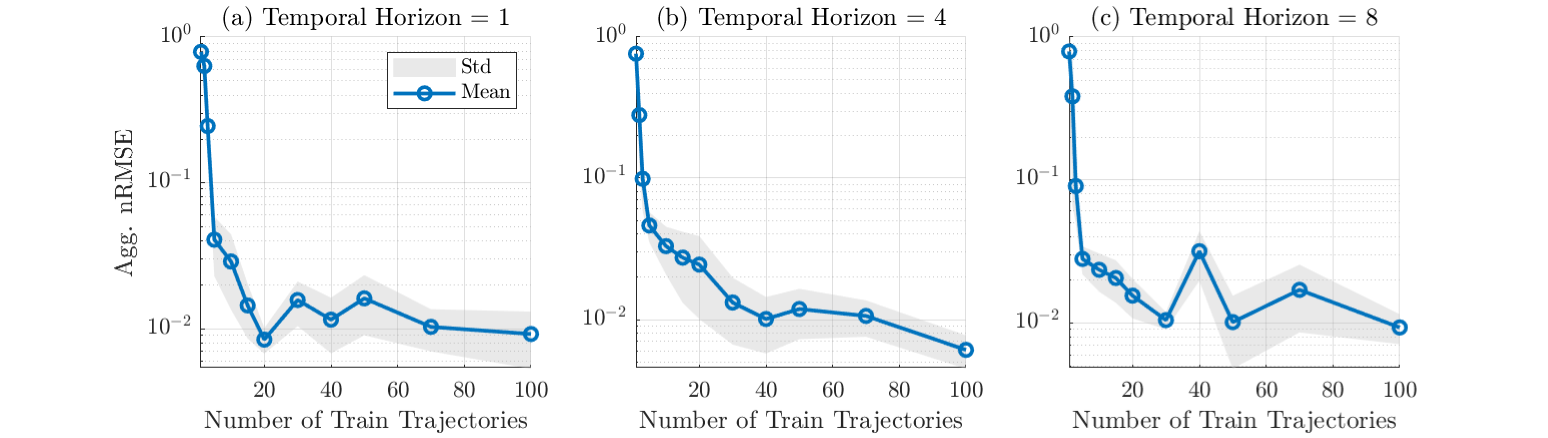}}
  \caption{Effect of training trajectory number and temporal horizon on emulator generalization in the Lorenz system. The blue line indicates the aggregated nRMSE error across five trained networks, while the shaded region denotes the standard deviation.}
  \label{fig:range}
\end{figure}

First, we ablate both the number of trajectories and the length of the training temporal horizon. In this study, we kept the network architecture identical and evaluated combinations of temporal horizons of 100, 400, and 800 time steps together with training sets containing 1, 2, 3, 5, 10, 15, 20, 30, 40, 50, 70, and 100 trajectories. For each combination, we trained five neural networks and plotted a shaded line curve showing the mean aggregated nRMSE as a function of the number of dataset samples, with the shaded region indicating the corresponding standard deviation. The results are presented in Fig.~\ref{fig:range}.

Across all combinations of these two parameters, the performance improves consistently as the amount of training data increases, up to a certain number of training samples beyond which gains become minor. This indicates that the emulator’s temporal generalization capability can converge with substantially less data than that used in our setting. The 80,000‑snapshot dataset was chosen following the setup used in the prior study \citep{brunton2022data}, rather than representing a minimal requirement.

Owing to the chaotic nature of the Lorenz system, extending the temporal horizon effectively provides additional independent samples, which is reflected by the convergence of the three curves to a similar performance level. However, this does not suggest that the temporal horizon plays an insignificant role. For systems with multiple dynamical stages, an excessively short horizon may fail to include important physical regimes. For example, in the Burgers equation, a horizon that is too short may miss the shock propagation stage. This is relevant because data-driven emulators may struggle to recover such behavior unless it is sufficiently represented in the training data.

More generally, the amount of data required by our approach is not a universal number, but depends on several factors: (i) the dimensionality of the system, (ii) the degree of nonlinearity, (iii) whether the system has multiple stages, and (iv) how well the data cover the neighborhood of the base state at which the Jacobian is queried.

\subsection{Dataset regime}

To assess how the learned neural emulator generalizes beyond the region covered by the training data, we perform an ablation study by deliberately restricting the training trajectories to a single lobe of the Lorenz attractor. The purpose of this experiment is to examine how the accuracy of the surrogate-based Jacobian depends on the local support of the training data in state space.

\begin{figure}
\centering
\includegraphics[width=\textwidth]{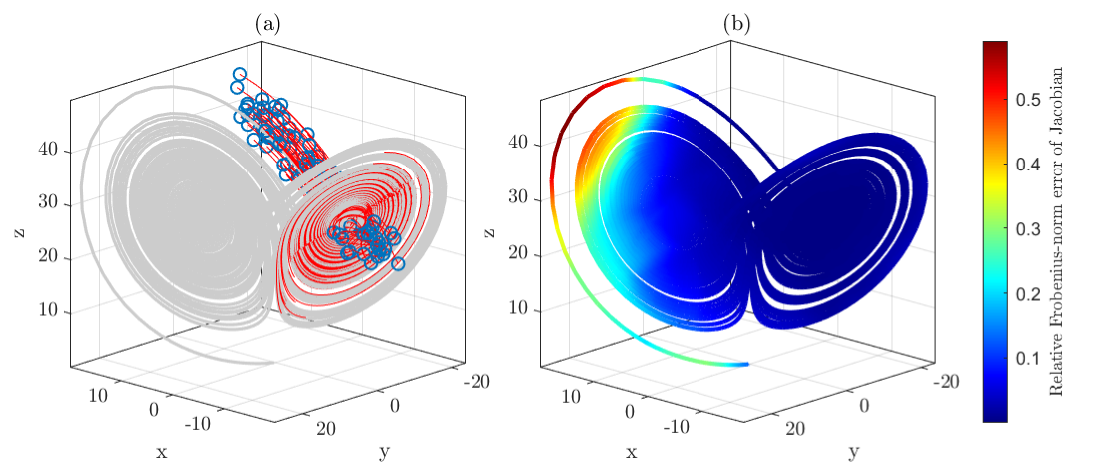}
\caption{
Effect of training regimes on emulator generalization in the Lorenz system. (a) Phase-space visualization of the Lorenz attractor, with the training trajectories (red curves) confined to one lobe and initial points marked (blue circles). The full attractors (gray curves) are generated by integrating trajectories initialized near the three equilibria. (b) Frobenius norm error between NN-based and operator-based Jacobians along the full attractors.
}
\label{fig:regime}
\end{figure}

Fig.~\ref{fig:regime}(a) illustrates the data construction. The 100 initial conditions denoted by blue circles are used to generate the training data, and the red curves denote the corresponding trajectories used for training. Each training trajectory is evolved for $8$ time units so that the total number of samples is the same as in the setting of the main text. To restrict the training data to the right lobe, a trajectory is accepted into the training set only if the proportion of points satisfying $x>0$ is less than $1\%$. Using this criterion, the associated trajectories are all concentrated on a single attractor lobe.

The gray background trajectory in Fig.~\ref{fig:regime}(a) is a long reference trajectory generated from three initial conditions located near the three Lorenz equilibria. Each trajectory is integrated with a time step $\Delta t = 0.01$ for a total duration of $50$ time units, so that the resulting reference set spans both lobes of the attractor as well as the transitions between them. After training, the neural emulator is evaluated via the relative Frobenius-norm error on these long reference trajectories.

We again train five neural networks with the same architecture as before. The color shown in Fig.~\ref{fig:regime}(b) corresponds to the mean relative Frobenius error evaluated along the reference trajectory. The results reveal a clear spatial dependence of the Jacobian accuracy. In the region of state space well represented by the training trajectories, the neural emulator recovers the Jacobian with substantially smaller error. As the evaluation points move away from that region, particularly toward the opposite lobe and the connecting transition region, the error increases progressively. This behavior indicates that the learned surrogate does not provide uniformly accurate local linearizations throughout the full attractor when its training support is restricted to only one portion of the state space.

This experiment highlights an important aspect of the proposed framework. Although the neural emulator can exhibit a certain degree of extrapolative capability, its Jacobian is most reliable in regions that are sufficiently sampled during training. In other words, the method should be interpreted as a tool for extracting local linear characteristics within a well-covered dynamical regime, rather than as a model expected to maintain quantitative Jacobian accuracy across dynamically distinct regimes that are absent from the training data.

\subsection{Noise level}

In this section, we performed an additional ablation study to assess the noise robustness of the neural emulator. This experiment probes the extent to which the learned one-step map and its induced local linearization remain reliable when the training trajectories are contaminated by measurement-like perturbations.

\begin{sidewaystable}
\centering
\begin{tabular}{cccccc} \toprule
\multirow{2}{*}{Base state $\boldsymbol{q}_b$} & \multicolumn{4}{c}{Eigenvalues} \\ \cline{2-6} \addlinespace[2pt]
& $1\%$ & $2\%$ & $5\%$ & $10\%$ & Ground Truth                                                                              \\ \midrule \addlinespace[5pt]
                             $\begin{bmatrix} 0 \\ 0 \\ 0 \end{bmatrix}$ & $\begin{bmatrix}1.1257\\0.9735\\0.7939\end{bmatrix}$ & $\begin{bmatrix}1.1256\\0.9735\\0.7979\end{bmatrix}$ & $\begin{bmatrix}1.1254\\0.9811\\0.7600\end{bmatrix}$ & $\begin{bmatrix}1.1243\\1.0135\\0.6988\end{bmatrix}$ &
                             $\begin{bmatrix}1.1256\\0.9737\\0.7959\end{bmatrix}$\\ \addlinespace[5pt]
 $\begin{bmatrix} 6\sqrt{2} \\ 6\sqrt{2} \\ 27 \end{bmatrix}$ & $\begin{bmatrix}0.8685\\0.9957\pm0.1018i\end{bmatrix}$              & $\begin{bmatrix}0.8590\\0.9947\pm0.1018i\end{bmatrix}$              & $\begin{bmatrix}0.8190\\0.9910\pm0.0997i\end{bmatrix}$              & $\begin{bmatrix}0.6206\\0.9553\pm0.0958i\end{bmatrix}$              &
 $\begin{bmatrix}0.8706\\0.9957\pm0.1019i\end{bmatrix}$              \\ \addlinespace[5pt]
                             $\begin{bmatrix} -6\sqrt{2} \\ -6\sqrt{2} \\ 27 \end{bmatrix}$ & $\begin{bmatrix}0.8678\\0.9958\pm0.1014i\end{bmatrix}$              & $\begin{bmatrix}0.8615\\0.9948\pm0.1014i\end{bmatrix}$              & $\begin{bmatrix}0.8190\\0.9877\pm0.0994i\end{bmatrix}$              & $\begin{bmatrix}0.6313\\0.9481\pm0.0868i\end{bmatrix}$  
                                    &
                             $\begin{bmatrix}0.8706\\0.9957\pm0.1019i\end{bmatrix}$  
                                    \\ \addlinespace[5pt] \midrule \addlinespace[5pt]
$\begin{bmatrix} 1.4689 \\ 8.7045 \\ -7.6905 \end{bmatrix}$ & $\begin{bmatrix}1.1443\\0.9793\\0.7775\end{bmatrix}$ & $\begin{bmatrix}1.1474\\0.9571\\0.7881\end{bmatrix}$ & $\begin{bmatrix}1.1452\\0.9820\\0.7661\end{bmatrix}$ & $\begin{bmatrix}1.1609\\0.9341\\0.7780\end{bmatrix}$ &
$\begin{bmatrix}1.1461\\0.9775\\0.7786\end{bmatrix}$ \\ \addlinespace[5pt] \bottomrule
\end{tabular}
\caption{Comparison of Jacobian eigenspectra obtained from neural emulators trained on the Lorenz dataset with different levels of noise: $1\%$, $2\%$, $5\%$, and $10\%$.}
\label{tab:noise}
\end{sidewaystable}

We perturbed the training data with additive random noise with noise levels of $1\%$, $2\%$, $5\%$, and $10\%$ and then constructed the corresponding one-step state-transition pairs from the noisy data. For each noise level, to evaluate the fidelity of the learned local dynamics, we compared the eigenvalues of the Jacobians predicted by the neural emulator at the three Lorenz equilibria and one non-equilibrium state against the analytical reference values; the results are summarized in Tab.~\ref{tab:noise}.

The results show that the emulator remains remarkably robust to moderate noise contamination in the training data. As the noise level increases from $1\%$ to $10\%$, the predicted eigenvalues indeed deviate from the ground truth progressively, indicating the expected degradation in local linear characteristic accuracy. 

As evidenced in Tab.~\ref{tab:noise}, this deviation is acceptable within low to moderate noise levels, depending on the stability properties of the evaluated point. The spectral properties of two fixed points associated with the attractor centers are consistently more difficult to predict than the fixed point at the origin and the non-equilibrium state. This increased difficulty is likely related to the strongly chaotic dynamics near the attractor centers in the Lorenz system, where trajectories initiated from nearby points can diverge at different rates and along different directions.

\begin{figure}
\centering
\includegraphics[width=0.8\linewidth]{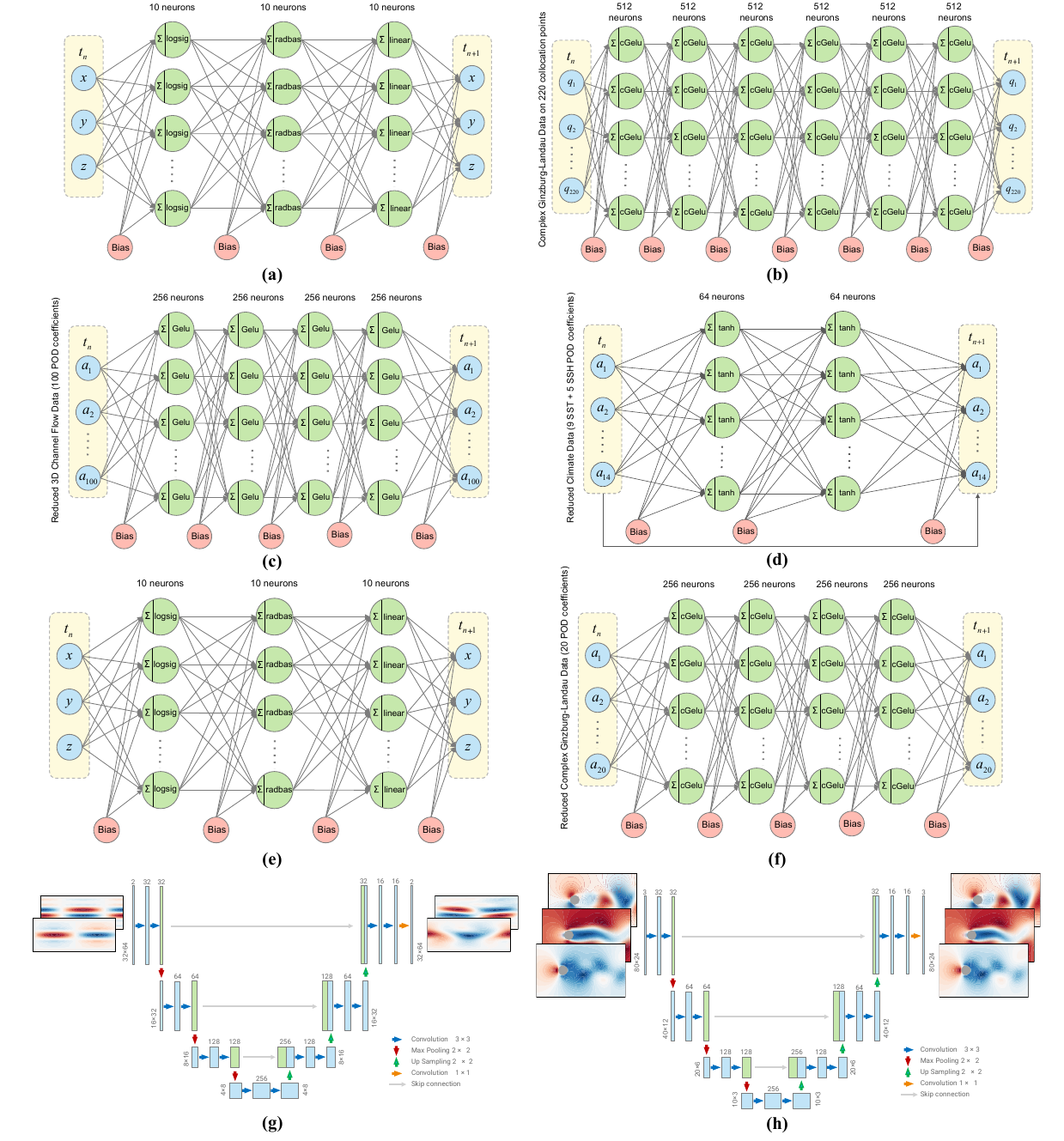}
\caption{Neural emulator architecture for all examples used in this paper. (a) A three-layer perceptron with heterogeneous activation functions (log-sigmoid, radial basis, and linear), used for the mean-field model of cylinder flow. (b) A six-layer complex-valued perceptron with complex Gelu activation function \citep{hendrycks2016gaussian} for each layer, used for the full-order complex Ginzburg-Landau system. (c) A four-layer perceptron used for the POD-reduced 3D channel flow system. (d) A two-layer residual perceptron used for the POD-reduced cyclostationary climate system. (e) A three-layer perceptron used for the Lorenz system. (f) A four-layer complex-valued perceptron used for the POD-reduced complex Ginzburg-Landau system. (g) A classical U-net used for the 2D channel flow system, where the two-channel inputs and outputs are the perturbation fields of streamwise and transverse velocity around the base flow, respectively.  (h) A classical U-net used for the 2D cylinder flow system, where the three-channel inputs and outputs are the pressure, streamwise velocity, and transverse velocity fields, respectively.}\label{fig:nn}
\end{figure}

\section*{Data Availability}
The datasets and source code generated during the current study are now available at \href{https://github.com/tum-pbs/NonlinearRA}{https://github.com/tum-pbs/NonlinearRA}. The previously published platform \citep{koehler2024apebench} used in this work is available at \href{https://github.com/tum-pbs/apebench}{https://github.com/tum-pbs/apebench}. Previously published data were used for this work \href{https://doi.org/10.1029/2002JD002670}{https://doi.org/10.1029/2002JD002670} \citep{rayner2003global} and \href{https://doi.org/10.1002/qj.2063}{https://doi.org/10.1002/qj.2063} \citep{balmaseda2013evaluation}.

\section*{Acknowledgements}
C.W. is supported by the China Scholarship Council (No. 202406340048) and L.C. is supported by the National Natural Science Foundation of China (Grant No. 92470120). We are grateful for many fruitful discussions with Jean-Christophe Loiseau and Eduardo Martini. We would also like to thank Benjamin Herrmann for kindly providing the dataset used in his paper.

\section*{Competing Interests}
The authors declare no competing interests.

\newpage
\bibliographystyle{unsrt} 
\bibliography{nc-sample}

\end{document}